\documentclass[12pt,twoside,a4paper]{article}
\usepackage{amsmath, amsthm, amssymb}

\usepackage[dvips]{epsfig}
\newcommand{\agt}{\,\rlap{\lower 3.5 pt \hbox{$\mathchar \sim$}} \raise 1pt
 \hbox {$>$}\,}
\newcommand{\alt}{\,\rlap{\lower 3.5 pt \hbox{$\mathchar \sim$}} \raise 1pt
 \hbox {$<$}\,}
\newcommand{\li}{\mathop{\mathrm{Li}_2}\nolimits}
\newcommand{\diff}{{\rm d}}

\catcode`@=11
\newcount\@tempcntc
\def\@citex[#1]#2{\if@filesw\immediate\write\@auxout{\string\citation{#2}}\fi
  \@tempcnta\z@\@tempcntb\m@ne\def\@citea{}\@cite{\@for\@citeb:=#2\do
    {\@ifundefined
       {b@\@citeb}{\@citeo\@tempcntb\m@ne\@citea\def\@citea{,}{\bf ?}\@warning
       {Citation `\@citeb' on page \thepage \space undefined}}%
    {\setbox\z@\hbox{\global\@tempcntc0\csname b@\@citeb\endcsname\relax}%
     \ifnum\@tempcntc=\z@ \@citeo\@tempcntb\m@ne
       \@citea\def\@citea{,}\hbox{\csname b@\@citeb\endcsname}%
     \else
      \advance\@tempcntb\@ne
      \ifnum\@tempcntb=\@tempcntc
      \else\advance\@tempcntb\m@ne\@citeo
      \@tempcnta\@tempcntc\@tempcntb\@tempcntc\fi\fi}}\@citeo}{#1}}
\def\@citeo{\ifnum\@tempcnta>\@tempcntb\else\@citea\def\@citea{,}%
  \ifnum\@tempcnta=\@tempcntb\the\@tempcnta\else
   {\advance\@tempcnta\@ne\ifnum\@tempcnta=\@tempcntb \else \def\@citea{--}\fi
    \advance\@tempcnta\m@ne\the\@tempcnta\@citea\the\@tempcntb}\fi\fi}
\catcode`@=12

\begin{document}
\thispagestyle{empty} 
\title{
\vskip-3cm
{\baselineskip14pt
\centerline{\normalsize DESY~07-215 \hfill ISSN 0418-9833}
\centerline{\normalsize LPSC~07-131 \hfill}
\centerline{\normalsize arXiv:0712.0481 [hep-ph] \hfill}
\centerline{\normalsize December 2007 \hfill}} 
\vskip1.5cm
\boldmath
{\bf Charmed-Meson Fragmentation Functions}\\
{\bf with Finite-Mass Corrections}
\unboldmath}\author{T.~Kneesch$^1$, B.A.~Kniehl$^1$, G.~Kramer$^1$, and
I.~Schienbein$^2$\bigskip
\\
{\normalsize $^1$ II. Institut f\"ur Theoretische Physik, Universit\"at
Hamburg,}\\
\normalsize{Luruper Chaussee 149, 22761 Hamburg, Germany}\bigskip
\\
\normalsize{$^2$ Laboratoire de Physique Subatomique et de Cosmologie,
Universit\'e Joseph Fourier}\\
\normalsize{Grenoble 1, CNRS/IN2P3, Institut National
Polytechnique de Grenoble,}\\
\normalsize{53 avenue des Martyrs, 38026 Grenoble, France}}

\date{}
\maketitle
\begin{abstract}
We elaborate the inclusive production of single heavy-flavored hadrons in
$e^+e^-$ annihilation at next-to-leading order in the general-mass
variable-flavor-number scheme.
In this framework, we determine non-perturbative fragmentation functions for
$D^0$, $D^+$, and $D^{*+}$ mesons by fitting experimental data from the
Belle, CLEO, ALEPH, and OPAL Collaborations, taking dominant electroweak
corrections due to photonic initial-state radiation into account.
We assess the significance of finite-mass effects through comparisons with a
similar analysis in the zero-mass variable-flavor-number scheme.
Under Belle and CLEO experimental conditions, charmed-hadron mass effects on
the phase space turn out to be appreciable, while charm-quark mass effects on
the partonic matrix elements are less important.

\medskip

\end{abstract}

\clearpage

\section{Introduction}
\label{sec:one}

In previous work \cite{Kniehl:2006mw}, two of us determined non-perturbative 
$D^0$, $D^+$, $D^{*+}$, $D_s^+$, and $\Lambda_c^+$ fragmentation functions
(FFs), both at leading order (LO) and next-to-leading order (NLO) in the 
modified minimal-subtraction ($\overline{\mathrm{MS}}$) factorization scheme,
by fitting the fractional-energy spectra of these hadrons measured by the OPAL
Collaboration \cite{Alexander:1996wy,Ackerstaff:1997ki} in $e^+e^-$
annihilation on the $Z$-boson resonance at the CERN Large Electron-Positron
Collider (LEP1).
Apart from untagged cross sections, they also measured the contributions
arising from $Z \rightarrow b\overline{b}$ decays.
This enabled the authors of Ref.~\cite{Kniehl:2006mw} to obtain specific FFs
for the transitions $c,b\to D^0,D^+,D^{*+},D_s^+,\Lambda_c^+$.
The strategy adopted in Ref.~\cite{Kniehl:2006mw} was very similar to the one
underlying Ref.~\cite{Binnewies:1997gz}, in which also ALEPH data
\cite{Barate:1999bg} were fitted, and Ref.~\cite{Kniehl:2005de}.
The FFs obtained in Ref.~\cite{Kniehl:2006mw} were used as input for a NLO
study \cite{Kniehl:2005ej} of charmed-meson hadroproduction in
$p\overline{p}$ collisions, which yielded reasonable agreement with data
collected by the CDF Collaboration in run II at the Tevatron
\cite{Acosta:2003ax}.

Recently, new data on charmed-meson production with much higher accuracy have
been presented by the Belle Collaboration \cite{Seuster:2005tr} at the KEK
Asymmetric Electron-Positron Collider for $B$ Physics (KEKB) and the CLEO
Collaboration \cite{Artuso:2004pj} at the Cornell Electron-Positron Storage
Ring (CESR). 
These data offer us the possibility to determine the non-perturbative initial
conditions of the FFs much more accurately.
Furthermore, the large span in center-of-mass (c.m.) energy ($\sqrt{s}$)
ranging from 10.5~GeV \cite{Seuster:2005tr,Artuso:2004pj} way up to 91.2~GeV
\cite{Alexander:1996wy,Ackerstaff:1997ki,Barate:1999bg} provides us with a
powerful lever arm to test the
Dokshitzer-Gribov-Lipatov-Altarelli-Parisi (DGLAP) \cite{Gribov:1972ri}
evolution of the FFs.
These new FFs will enable us to improve our theoretical predictions for the
charmed-meson hadroproduction cross sections
\cite{Kniehl:2005ej,Kniehl:2004fy} to be compared with the CDF data
\cite{Acosta:2003ax}.

The data from Belle and CLEO are located much closer to the thresholds
$\sqrt{s}=2m_c$ and $\sqrt{s}=2m_b$ of the transitions $c \rightarrow H_c$ and
$b \rightarrow H_b$, where $H_c$ and $H_b$ stand for generic $c$ or $b$
hadrons, respectively, than those from ALEPH and OPAL.
It might thus be a questionable approximation to treat the partonic cross
sections for $e^+e^- \to c+X$ and $e^+e^-\to b+X$ in the massless
approximation, with $m_c=m_b=0$, as was done in
Refs.~\cite{Kniehl:2006mw,Binnewies:1997gz,Kniehl:2005de}, where LEP1 data
were fitted.
Therefore, we take into account the finite quark mass corrections of the form 
$m^2/s$ ($m=m_c,m_b$) in the partonic cross sections to test their
significance.
Similar studies based on perturbative FFs \cite{Mele:1990cw} may be found in
Refs.~\cite{Cacciari:2005uk,Corcella:2007tg}.

The outline of this paper is as follows.
In Section~\ref{sec:two}, we describe the theoretical formalism of
single-hadron inclusive production in $e^+e^-$ annihilation.
After reviewing the massless case in Section~\ref{sec:twoa}, we explain how to
include the full mass corrections in Section~\ref{sec:twob}.
Lengthy expressions are relegated to Appendices~\ref{sec:appa} and
\ref{sec:appb}.
Specifically, we list the electroweak quark charges in Appendix~\ref{sec:appa}
and the mass-dependent coefficient functions in Appendix~\ref{sec:appb}.
In Section~\ref{sec:twoc}, we consider the electromagnetic initial-state
radiation (ISR) that is inherent to the Belle and CLEO data and explain how to
efficiently accommodate the ISR corrections in our fits.
In Section~\ref{sec:three}, we present several alternative FF sets.
They are obtained from global fits to Belle \cite{Seuster:2005tr}, CLEO
\cite{Artuso:2004pj}, ALEPH \cite{Barate:1999bg}, and OPAL
\cite{Alexander:1996wy,Ackerstaff:1997ki} data, and from separate fits to the
$B$-factory (Belle plus CLEO) and $Z$-factory (ALEPH plus OPAL) data, both for
$m\neq0$ and for $m=0$.
Finally, in Section~\ref{sec:four}, we present a summary and our conclusions.


\section{Formalism}
\label{sec:two}

We study the inclusive production of a single charmed hadron $H_c$, with mass
$m_H$, in $e^+e^-$ annihilation via a virtual photon ($\gamma$) or $Z$ boson,
\begin{equation}
e^+ + e^- \to (\gamma,Z) \to H_c + X,
\label{eq:eeprocess}
\end{equation}
where $X$ stands for the residual final state, which goes unobserved.
Specifically, we concentrate on the cases $H_c=D^0,D^+,D^{*+}$.
In the following, we explain how to calculate the cross section of
process~(\ref{eq:eeprocess}) at NLO in the parton model of QCD, both in the
zero-mass (ZM) approach, where all quark masses are neglected, and in
the general-mass (GM) approach, where the $c$ and $b$ quarks are taken to be
massive.
We denote the four-momenta of the virtual gauge boson and the $H_c$ hadron by
$q$ and $p_H$, respectively, so that $s=q^2$ and $m_H^2=p_H^2$, and introduce
the scaling variable $x=2(p_H\cdot q)/q^2$.
We call the energy of $H_c$ and the angle of its three-momentum w.r.t.\ the
beam axis in the c.m.\ frame $E$ and $\theta$, respectively.
Then, $x=2E/\sqrt{s}$ measures the energy of $H_c$ in units of the beam energy.

For unpolarized beams and observed hadrons, the cross section of
process~(\ref{eq:eeprocess}) at a given value of $\sqrt{s}$ can only depend on
$E$ and $\theta$.
Since the virtual boson has spin one, the most general form of the
differential cross section then reads
\begin{equation}
\frac{\diff^2 \sigma}{\diff x \, \diff \cos \theta}
= \frac{3}{8}(1+\cos^2 \theta) \frac{\mathrm{d}\sigma^T}{\diff x} 
+ \frac{3}{4}\sin^2 \theta \frac{\diff \sigma^L}{\diff x} 
+ \frac{3}{4} \cos \theta \frac{\diff \sigma^A}{\diff x}.
\label{eq:xstheta}
\end{equation}
The three terms on the right-hand side are the transverse, longitudinal,
and asymmetric contributions, respectively.
The first two are associated with the corresponding polarization states of the
virtual boson with respect to the direction of the observed hadron.
The asymmetric contribution is due to the parity-violating interference terms
and is not present in QED.
The transverse and longitudinal parts are normalized so that
\begin{equation}
\frac{\diff \sigma}{\diff x} = \int\limits^{+1}_{-1} \diff \cos \theta 
\frac{\diff^2 \sigma}{\diff x\, \diff \cos \theta}  
= \frac{\diff \sigma^T}{\diff x} + \frac{\diff \sigma^L}{\diff x}.
\label{eq:xs}
\end{equation}

In the parton model, each component $\diff\sigma^P/\diff x$ ($P=T,L$) on the
right-hand side of Eq.~(\ref{eq:xs}) can be written, up to power corrections,
as a sum of convolutions of partonic cross sections
$\diff\sigma_a^P(y,\mu,\mu_f)/\diff y$, where
$a=g,u,\overline{u},\ldots,b,\overline{b}$ is the fragmenting parton with
four-momentum $p_a$, $y=2(p_a\cdot q)/q^2$, and $\mu$ and $\mu_f$ are the
renormalization and factorization scales, respectively, with FFs
$D_a(z,\mu_f)$, where $z=x/y$ is the fraction of energy passed on from parton
$a$ to hadron $H_c$ in the c.m.\ frame, as
\begin{equation}
\frac{\diff \sigma^P}{\diff x}(x,s)=\sum\limits_a
\int\limits_{y_\mathrm{min}}^{y_\mathrm{max}} \frac{\diff y}{y}\,
\frac{\diff \sigma_a^P}{\diff y}(y,\mu,\mu_f)D_a\left(\frac{x}{y},\mu_f\right),
\label{eq:convolution}
\end{equation}
where the values of $y_\mathrm{min}$ and $y_\mathrm{max}$ are subject to mass
effects to be discussed below. 
At NLO, $\mu_f$ defines the scale, where the divergence associated with
collinear gluon radiation off a massless primary quark or antiquark is to be
subtracted.


\subsection{ZM approach}
\label{sec:twoa}

At NLO in the $\overline{\mathrm{MS}}$ scheme, the cross sections of the
relevant partonic subprocesses are given by \cite{Baier:1979sp}
\begin{eqnarray}
\frac{\diff \sigma_{q_i}}{\diff y}(y,\mu,\mu_f)
&=&N_c\sigma_0\left(V_{q_i}^2+A_{q_i}^2\right)
\left\{ \delta(1-y) + \frac{\alpha_s(\mu)}{2\pi}
\left[ P_{q\to q}^{(0,T)}(y) \ln \frac{s}{\mu_f^2} + C_q(y) \right] \right\},
\nonumber\\
\frac{\diff \sigma_{g}}{\diff y}(y,\mu,\mu_f)
&=&2N_c\sigma_0\sum\limits_{i=1}^{n_f}\left(V_{q_i}^2+A_{q_i}^2\right)
\frac{\alpha_s(\mu)}{2\pi}\left[P_{q\to g}^{(0,T)}(y)\ln\frac{s}{\mu_f^2}
+C_g(y) \right].
\label{eq:ZM-XS}
\end{eqnarray}
Here, $N_c=3$ is the number of quark colors; 
\begin{equation}
\sigma_0=\frac{4\pi\alpha^2}{3s},
\label{eq:sig0}
\end{equation}
with $\alpha$ being Sommerfeld's fine-structure constant, is the total cross
section of $e^+e^-\to\mu^+\mu^-$ for massless muons;
$V_{q_i}$ and $A_{q_i}$ are the effective vector and axial-vector couplings of
quark $q_i$ to the photon and the $Z$ boson including propagator adjustments,
which are listed in Appendix~\ref{sec:appa};
$P_{a\to b}^{(0,T)}$ are the LO timelike splitting functions
\cite{Gribov:1972ri},
\begin{eqnarray}
P_{q\to q}^{(0,T)}(y)&=&C_F\left[\frac{3}{2}\delta(1-y)
+\frac{1+y^2}{(1-y)_+}\right],
\nonumber\\
P_{q\to g}^{(0,T)}(y)&=&C_F\frac{1+(1-y)^2}{y};
\end{eqnarray}
and the coefficient functions read \cite{Baier:1979sp}
\begin{eqnarray}
C_q(y) & = & C_F \left\{ \left( -\frac{9}{2} + \frac{2}{3}\pi^2 \right) 
\delta(1-y) - \frac{3}{2} \left( \frac{1}{1-y} \right)_+ 
+ 2\left[ \frac{\ln (1-y)}{1-y} \right]_+ + \frac{5}{2} - \frac{3}{2}y
\right.\nonumber\\ 
& &{}+\left. 4 \frac{\ln y}{1-y} - (1+y) [2 \ln y + \ln(1-y)] \right\},
\nonumber\\
C_g(y) & = & C_F \frac{1+(1-y)^2}{y} [2 \ln y + \ln (1-y)],
\label{eq:ZM-XS-Cfunc}
\end{eqnarray}
where $C_F=(N_c^2-1)/(2N_c)=4/3$ and the plus distributions are defined as
usual. 
We evaluate the strong-coupling constant $\alpha_s$ using the two-loop
formula with $n_f=5$ quark flavors.
We identify $\mu=\mu_f=\sqrt{s}$ , so that in Eq.~(\ref{eq:ZM-XS}) the
terms proportional to $\ln (s/\mu_f^2)$ vanish.
We observe that $C_g(y)<0$ for any value of $y$, so that the gluon contributes
destructively to Eq.~(\ref{eq:ZM-XS}). 
The bounds of integration in Eq.~(\ref{eq:convolution}) are $y_\mathrm{min}=x$,
$y_\mathrm{max}=1$, and we have $\sqrt{\rho_H}\le x\le1$, where
$\rho_H=4m_H^2/s$.

For later use, we also list the total hadronic cross section at NLO,
\begin{equation}
\sigma_\mathrm{tot}=N_c\sigma_0\sum\limits_{i=1}^{n_f}
\left(V_{q_i}^2+A_{q_i}^2\right)\left[1+\frac{\alpha_s(\mu)}{2\pi}C_F
\frac{3}{2}\right].
\label{eq:sigtot}
\end{equation}

The $x$ dependences of the FFs are not yet calculable from first principles.
However, once they are given at some initial fragmentation scale $\mu_0$,
their $\mu_f$ evolution is determined by the DGLAP evolution equations
\cite{Gribov:1972ri},
\begin{equation}
\frac{\diff}{\diff \ln \mu_f^2} D_a(x,\mu_f) 
= \frac{\alpha_s(\mu)}{2\pi} \sum\limits_b \int\limits_x^1
\frac{\diff y}{y} P_{a\to b}^T(y,\alpha_s(\mu))
D_b\left(\frac{x}{y},\mu_f\right).
\label{eq:DGLAP}
\end{equation}
Specifically, we use $\mu_0=m_c$ for
$a=g,u,\overline{u},d,\overline{d},s,\overline{s},c,\overline{c}$ and
$\mu_0=m_b$ for $a=b,\overline{b}$.
Our task is thus to construct a model for the $z$ dependences of
$D_a(z,\mu_0)$, which upon evolution to $\mu_f=\sqrt{s}$ fit the data at that
c.m.\ energy.

The above formalism is identical to the one that is routinely used in the
literature for the inclusive production of single light hadrons
\cite{Binnewies:1994ju}.
The non-zero values of the $c$- and $b$-quark masses only enter through the
initial conditions of the FFs, and the mass of the heavy hadron sets the lower
bound on the scaling variable $x$.


\subsection{GM approach}
\label{sec:twob}

We derived the partonic cross sections at NLO for non-zero quark masses
adopting the on-shell definition of the latter and found agreement with
Ref.~\cite{Nason:1993xx}.
We take the pole masses of the $c$ and $b$ quarks to be $m_c=1.5$~GeV and
$m_b=5.0$~GeV, respectively.
The finite-mass corrections are generally of order $m^2/s$ ($m=m_c,m_b$). 
They can be sizeable for sufficiently small values of $\sqrt{s}$.
In our study, where the smallest value of $\sqrt{s}$ is 10.52~GeV, they
reach at least 2\% for the $c$ quark, but might be much larger depending on
the coefficient of $m^2/s$ in the LO cross section.
On the other hand, they are quite substantial for the $b$ quark at values of
$\sqrt{s}$ just above the $b\overline{b}$ production threshold.

In the following, we consider finite-mass effects of order $m^2/s$ only in the
production dynamics, but not in the decays of the produced $H_c$ mesons,
which are affected by kinematic power corrections.
This would be beyond the scope of our analysis and is not required because
the experimental analyses actually provide production cross sections.

In order to expose the connection with the ZM limit and to establish the
finite subtraction terms, which are needed for the evaluation of the GM cross
section in the $\overline{\mathrm{MS}}$ factorization scheme, we report the
relevant cross section formulae for the GM case.
We largely adopt the notation of Ref.~\cite{Nason:1993xx}.

At NLO, there is a contribution from real gluon radiation.
The gluon can either be included in the hadronic system $X$ or it can act as
the leading parton initiating a hadron jet that contains the $H_c$ meson.
Therefore, we also need to consider at the parton level single-gluon inclusive
production with $X$ including a $c\overline{c}$ or $b\overline{b}$ pair.

The partonic cross sections of single-heavy-quark inclusive production read
\begin{equation}
\frac{\diff \sigma_{q_i}^P}{\diff y}(y,\mu) 
=N_c\sigma_0\left[V_{q_i}^2F_P^{(v)}(y,\rho)+A_{q_i}^2 F_P^{(a)}(y,\rho)
\right],
\label{eq:gm}
\end{equation}
where $\rho=4m^2/s$ with $m=m_c,m_b$.
Notice that the vector and axial-vector contributions differ by finite-mass
terms.
At NLO, the coefficient functions $F_P^{(u)}(y,\rho)$, with $P=T,L$ and
$u=v,a$, in Eq.~(\ref{eq:gm}) may be decomposed as \cite{Nason:1993xx}:
\begin{equation}
F_P^{(u)}(y,\rho)=\delta(1-y)B_P^{(u)}(\rho)+\frac{\alpha_s(\mu)}{2\pi}
\left[\delta(1-y)S_P^{(u)}(\rho)+\left(\frac{1}{1-y}\right)_+
R_P^{(u)}(y,\rho)\right].
\label{eq:XS-NLODecomposition}
\end{equation}
The LO terms read
\begin{equation}
B_T^{(v)}(\rho)=\beta,\qquad
B_L^{(v)}(\rho)=\frac{\rho\beta}{2},\qquad
B_T^{(a)}(\rho)=\beta^3,\qquad
B_L^{(a)}(\rho)=0,
\end{equation}
where $\beta=\sqrt{1-\rho}$.
In the massless limit, we have $B_T^{(v)}(\rho)=B_T^{(a)}(\rho)=1$ and
$B_L^{(v)}(\rho)=B_L^{(a)}(\rho)=0$.
The NLO terms, $S_P^{(u)}(\rho)$ and $R_P^{(u)}(y,\rho)$, may be found in
Appendix~\ref{sec:appb}.
The partonic cross section of single-gluon inclusive production reads
\begin{equation}
\frac{\diff \sigma_g^P}{\diff y}(y,\mu) 
=N_c\sigma_0\frac{\alpha_s(\mu)}{2\pi}\,
\left[V_{q_i}^2 G_P^{(v)}(y,\rho) + A_{q_i}^2 G_P^{(a)}(y,\rho)\right],
\label{eq:gmg}
\end{equation}
with the coefficient functions $G_P^{(u)}(y,\rho)$ given in
Appendix~\ref{sec:appb}.
Deviating from the notation of Ref.~\cite{Nason:1993xx}, we included the
factor $C_F$ in the functions $S_P^{(u)}(\rho)$, $R_P^{(u)}(y,\rho)$, and
$G_P^{(u)}(y,\rho)$ to simplify the comparison with the ZM approximation. 
Through this comparison, we recover the so-called perturbative FFs
\cite{Mele:1990cw}, from which the subtraction terms for the conversion from
the NLO calculation with finite quark masses in the on-shell scheme
\cite{Nason:1993xx} to the GM variable-flavor-number scheme
\cite{Kniehl:2005ej,Kniehl:2004fy} are constructed.
The result thus obtained contains all the finite-mass terms and, at the same
time, smoothly approaches the ZM result in the limit $m\to0$. 

Thanks to our specific choice of hadronic and partonic scaling variables, $x$
and $y$, respectively, Eq.~(\ref{eq:convolution}) applies to the GM case as it
stands.
The bounds of integration in $y$ and the allowed $x$ range now depend on the
partonic subprocess and on the fragmenting parton $a$.
In the case of heavy-quark fragmentation, we have
$y_\mathrm{min}=\max(x,\sqrt{\rho})$, $y_\mathrm{max}=1$, and
$\sqrt{\rho_H}\le x\le1$.
If a gluon fragments and there is also a heavy-quark pair in the final state,
then we have $y_\mathrm{min}=x$, $y_\mathrm{max}=\beta^2$, and
$\sqrt{\rho_H}\le x\le\beta^2$.
If there are only massless partons in the final state, then we have
$y_\mathrm{min}=x$, $y_\mathrm{max}=1$, and $\sqrt{\rho_H}\le x\le1$ as in
Section~\ref{sec:twoa}.

The experimental data at $\sqrt{s}=m_Z$, collected by OPAL 
\cite{Alexander:1996wy,Ackerstaff:1997ki} and ALEPH \cite{Barate:1999bg} at
LEP1, come in the form $\diff\sigma/\diff x$ as a function of the scaling
variable $x$ introduced above.
As explained above, the maximum $x$ range is $\sqrt{\rho_H}\le x\le1$.
On the other hand, Belle \cite{Seuster:2005tr} and CLEO \cite{Artuso:2004pj} 
present their data as distributions $\diff\sigma /\diff x_p$ in the scaled
momentum $x_p=p/p_\mathrm{max}=\sqrt{(x^2-\rho_H)/(1-\rho_H)}$, with allowed
values $0\le x_p\le1$.
The conversion formula reads
\begin{equation}
\frac{\diff\sigma}{\diff x_p}(x_p)=(1-\rho_H)\frac{x_p}{x}\,
\frac{\diff \sigma}{\diff x}(x),
\label{eq:xp}
\end{equation}
with $x= \sqrt{(1-\rho_H)x_p^2+\rho_H}$.

We conclude this section with a remark concerning parton model kinematics of
fragmentation in the presence of finite quark and hadron masses.
In the picture where the fragmenting parton $a$ creates a jet that includes
the observed hadron $H$, the parton virtuality $p_a^2$ must exceed the hadron 
mass square $m_H^2$.
In fact, if the scaling variable $z$ is defined in terms of light-cone
momenta, as $z=p_H^+/p_a^+=(p_H^0+p_H^3)/(p_a^0+p_a^3)$
\cite{Albino:2005gd}, we have $p_a^2>m_H^2/z$ \cite{Kniehl:1999xc}.
Since the outgoing partons are taken to be on-shell in the parton model, this
inequality becomes $m^2>m_H^2/z$.
In our case, this is only satisfied for the transition $b\to H_c$ in the
constrained $z$ range $m_H^2/m^2<z<1$.

At this point, we find find it instructive to insert a digression on the
massive kinematics of fragmentation, a topic which has received very little
attention in the literature.
Let us view the fragmentation process $b\to H_c$ as a decay $b\to H_c+X$,
where the invariant mass $m_X$ of the hadronic system $X$, which we may treat
as one effective particle, can be tuned in the range $0\le m_X\le m-m_H$.
Starting from the rest frame of the decay, we perform a Lorentz boost along the
three-momentum of $H_c$, such that the energy of $H_c$ becomes
$p_H^0=x\sqrt{s}/2$, where $\sqrt{s}$ is the $e^+e^-$ c.m.\ energy.
This corresponds to the situation where $H_c$ is emitted collinearly from $b$,
carrying energy $p_b^0=y\sqrt{s}/2$, in the laboratory frame.
For given values of $x$ and $m_X$, we have
\begin{equation}
y=\frac{1}{2m_H^2}\left[\left(m^2+m_H^2-m_X^2\right)x-
\sqrt{\lambda\left(m^2,m_H^2,m_X^2\right)(x^2-\rho_H)}\right],
\end{equation}
where $\lambda(a,b,c)=a^2+b^2+c^2-2(ab+bc+ca)$.
Owing to the constraint $\sqrt{\rho}\le y\le1$ on $y$ and the one on $m_X$
specified above, $x$ must lie in the range
$\sqrt{\rho_H}\le x\le x_\mathrm{max}$, with
\begin{equation}
x_\mathrm{max}=\frac{1}{2m^2}\left[m^2+m_H^2+\beta\left(m^2-m_H^2\right)
\right].
\end{equation}
For a given value of $x$, the allowed $y$ range is thus
$y_\mathrm{min}(x)\le y\le y_\mathrm{max}(x)$, where
\begin{eqnarray}
y_\mathrm{min}(x)&=&
\begin{cases}
\sqrt{\rho} & \text{if $\sqrt{\rho_H}\le x\le\frac{m^2+m_H^2}{m\sqrt{s}}$} \\
y_0 & \text{if $\frac{m^2+m_H^2}{m\sqrt{s}}< x\le x_\mathrm{max}$}
\end{cases},
\nonumber\\
y_\mathrm{max}(x)&=&
\begin{cases}
\min(y_0,1) & \text{if $\sqrt{\rho_H}\le x\le(m+m_H)\sqrt{\frac{m_H}{ms}}$} \\
\min\left(x\frac{m}{m_H},1\right) & 
\text{if $(m+m_H)\sqrt{\frac{m_H}{ms}}< x\le x_\mathrm{max}$}
\end{cases},
\end{eqnarray}
where
\begin{equation}
y_0=\frac{1}{2m_H^2}\left[\left(m^2+m_H^2\right)x-
\left(m^2-m_H^2\right)\sqrt{x^2-\rho_H}\right].
\end{equation}
For $\sqrt{s}\ge(m^2+m_H^2)/m_H$, the scaling variable $z=x/y$ appearing in
the $b\to H_c$ FF lies in the range
$2m_H^2/(m^2+m_H^2)\le z\le x_\mathrm{max}$. 

If $m_H\ll m$ is a good approximation, we may simplify these expressions by
putting $m_H=0$.
Then, we have $0\le x\le x_\mathrm{max}$, with
$x_\mathrm{max}=(1+\beta)/2$, and
\begin{eqnarray}
y_\mathrm{min}(x)&=&
\begin{cases}
\sqrt{\rho} & \text{if $0\le x\le\frac{\sqrt{\rho}}{2}$} \\
x+\frac{\rho}{4x} & \text{if $\frac{\sqrt{\rho}}{2}< x\le x_\mathrm{max}$}
\end{cases},
\nonumber\\
y_\mathrm{max}(x)&=&1.
\end{eqnarray}
Consequently, we have $0\le z\le x_\mathrm{max}$.
Finally, if also $m=0$, we recover the ZM situation,
$0\le x\le1$ and $x\le y\le1$, so that $0\le z\le1$.

\begin{figure}[ht]
\begin{center}
\epsfig{file=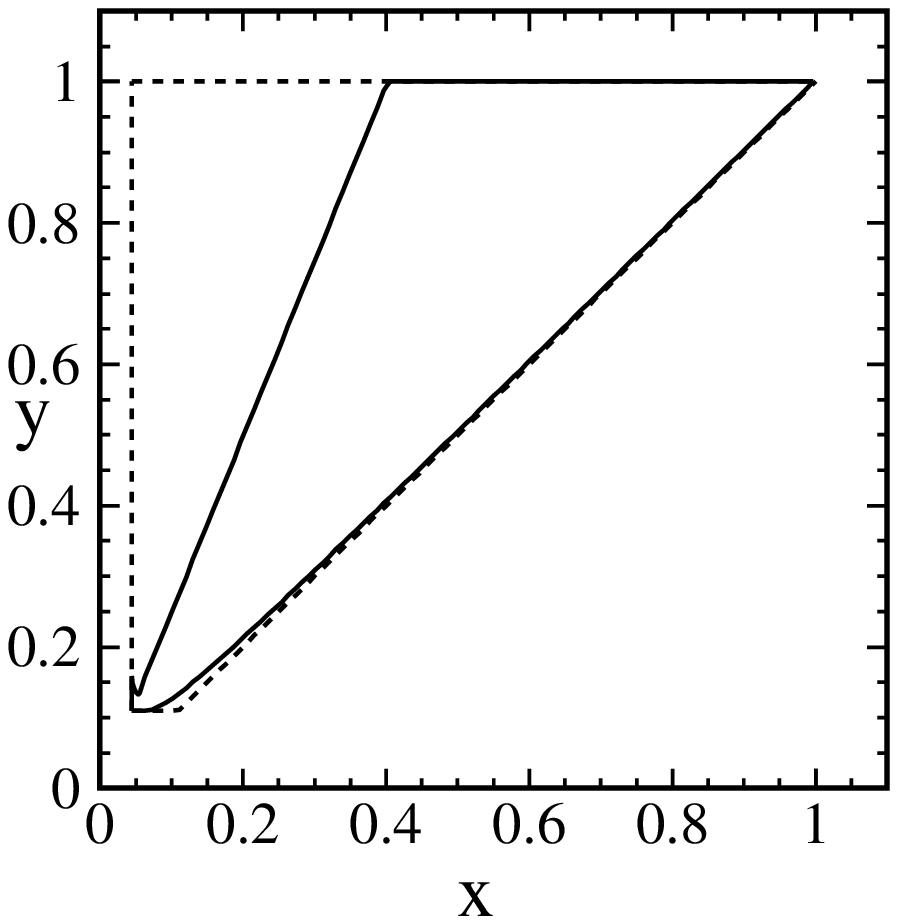,width=0.9\textwidth}
\end{center}
\caption{\label{Fig:kin}Kinematically accessible region in the $(x,y)$ plain
for $e^+e^-\to b+X$ and subsequent decay $b\to H_c+X$ with collinear emission
of $H_c$ assuming $\sqrt{s}=m_Z$, $m=5.0$~GeV, and $m_H=2.01$~GeV (solid line).
For comparison, also the region used in our analysis is shown (dashed line).} 
\end{figure}
In Fig.~\ref{Fig:kin}, the kinematically accessible region in the $(x,y)$
plain is shown for our fragmentation model of $b\to H_c$ assuming LEP1
experimental conditions, with $\sqrt{s}=m_Z$, $m=5.0$~GeV, and $m_H=2.01$~GeV
(solid line).
It does not exhaust the region encompassed by $\sqrt{\rho_H}\le x\le1$ and
$\max(x,\sqrt{\rho})\le y\le1$ that we adopt for the evaluation of
Eq.~(\ref{eq:convolution}) in the case of $b\to H_c$ (dashed line).
There is a wedge missing at small values of $x$ and large values of $y$,
{\it i.e.}\ small values of $z$.
The minimum value of $z$, $2m_H^2/(m^2+m_H^2)\approx0.28$, is reached at the
minimum value of $x$.
However, in the bulk of the missing wedge, we have $z\le m_H/m\approx0.40$.
As we shall see in Section~\ref{sec:three}, this excluded $z$ range actually
accommodates the peak of the $b\to H_c$ FF (see also
Table~\ref{Tab:AverageEnergy}).
Moreover, our simple fragmentation picture cannot describe the transitions
$a\to H_c$ for $a=u,d,s,c,g$.
For these reasons, we abandon it at this point.
A crucial conceptual drawback of the parton model applied to the inclusive
production of heavy hadrons is that fragmenting partons are taken to be
on-shell.
We escape this problem by imagining that the hadronic system initiated by the
fragmenting parton $a$ not only receives color, but also energy and momentum
from the rest of the event.
In this way, the accessible region in the $(x,y)$ plane is expanded to become
the one underlying our analysis.
By the same token, the masses of the fragmenting quark and the observed hadron
can be treated independently of each other, the former appearing in the
partonic cross sections and the latter in the hadronic phase space factor.
As explained above, we take $m=0<m_H$ in the ZM approach and $m,m_H>0$ in the
GM approach.
In Section~\ref{sec:three}, we also study the case $m=m_H=0$ for comparison.


\subsection{Electromagnetic initial-state radiation}
\label{sec:twoc}

The cross sections of inclusive single-hadron production in $e^+e^-$
annihilation measured by Belle \cite{Seuster:2005tr}, CLEO
\cite{Artuso:2004pj}, ALEPH \cite{Barate:1999bg}, and OPAL
\cite{Alexander:1996wy,Ackerstaff:1997ki} naturally include electroweak
corrections, which were not subtracted in the data analyses.
The bulk of these corrections is due to the effect of electromagnetic 
radiation emitted from the incoming electrons and positrons.
This ISR is suppressed by a factor of $\alpha$, but enhanced by the large
logarithm $\log(s/m_e^2)$, where $m_e$ is the electron mass. 
At Belle and CLEO energies, the hadronic cross section decreases with
increasing invariant mass of the hadronic system.
Since ISR reduces the hadronic mass, it leads to an increase in cross section.
The shape of the FF is also changed, since a fraction of the events takes
place at lower hadronic invariant mass.
The impact of ISR on the determination of FFs has already been considered in
Ref.~\cite{Cacciari:2005uk} and has been found to be non-negligible for the
analysis of the Belle and CLEO data.

The most straightforward way to correct for ISR would be to incorporate into
the general expression for the $e^+e^-\to H_c + X$ cross section without
photon radiation, presented in Section~\ref{sec:twoa}, the corrections due to
photon radiation off the initial-state leptons and use the resulting
expression to fit the FFs to the Belle and CLEO data.
This procedure would involve several additional numerical integrations in each
iteration of the fitting procedure and thus dramatically slow down the latter.
In the following, we explicitly derive an approximation formula, which reduces
the number of integrations to a manageable level and is still rather precise.
Our procedure differs from the one used in Ref.~\cite{Cacciari:2005uk}, where
the ISR corrections were subtracted in an iterative way from the experimental
data before the actual fit.
Our procedure is numerically more involved, but offers the advantage that, at
at the end, the ISR corrections precisely refer to the final $x$ distribution
resulting from the fit.

The dominant ISR corrections are conveniently incorporated using the
structure-function approach, in which the photon emission is taken to be
collinear to the incoming $e^\pm$ leptons
\cite{Kuraev:1985hb,Nicrosini:1986sm}.
In analogy to the factorization formula of the collinear parton model, the
ISR-corrected differential cross section $d\sigma_\mathrm{ISR}(p_+,p_-)$,
where $p_\pm$ denote the four-momenta of the incoming $e^\pm$ leptons, is
obtained by convoluting the uncorrected differential cross section
$d\sigma(p_+,p_-)$ with radiator functions $D_{e^\pm}(x_\pm,s)$, one for each
incoming leg, which measure the probabilities for the $e^\pm$ leptons to
retain the fractions $x_\pm$ of their energies after the emission of ISR, as
\begin{equation}
\diff\sigma_\mathrm{ISR}(p_+,p_-)=
\int\limits_0^1\diff x_+\int\limits_0^1\diff x_-\,
D_{e^+}(x_+,s)D_{e^-}(x_-,s)\diff\sigma(x_+p_+,x_-p_-).
\label{eq:fac}
\end{equation}
Using the method by Gribov and Lipatov \cite{Gribov:1972rt}, the leading
logarithms can be resummed to all orders, leading to the expression for
$D_{e^\pm}(x,s)$ in Eq.~(7) of Ref.~\cite{Nicrosini:1986sm}.
The structure-function approach was mostly applied to total cross sections in
the literature.
In this case, one integration can be carried out independently of the
considered process and leads to the luminosity function
\begin{equation}
H_{e^+e^-}(\tau,s)=\int\limits_0^1\diff x_+\int\limits_0^1\diff x_-
\delta(\tau-x_+x_-)D_{e^+}(x_+,s)D_{e^-}(x_-,s)
\label{eq:h}
\end{equation}
in $\tau=x_+x_-$.
The Gribov-Lipatov-resummed expression for it may be found in Eqs.~(8) and (9)
of Ref.~\cite{Nicrosini:1986sm}, in terms of the variable $\chi=1-\tau$.
However, if the cross section is differential w.r.t.\ variables, whose
differentials are not invariant under boosts along the beam axis, as in our
case, the situation is more involved \cite{Munehisa:1995si}.
In this case, we obtain from Eq.~(\ref{eq:fac}) the following master formula:
\begin{eqnarray}
\frac{\diff\sigma_\mathrm{ISR}}{\diff x}(x,s)
&=&\int\diff x_+\,\diff x_-\,\diff x^\prime\,\diff\cos\theta^\prime\,
\delta(x-x(x_+,x_-,x^\prime,\cos\theta^\prime))
D_{e^+}(x_+,s)D_{e^-}(x_-,s)
\nonumber\\
&&{}\times\frac{\diff^2\sigma}{\diff x^\prime\,\diff\cos\theta^\prime}
(x^\prime,\cos\theta^\prime,x_+x_-s),
\label{eq:master}
\end{eqnarray}
where the primed variables refer to the hadronic c.m.\ frame, which is reached
from the $e^+e^-$ c.m.\ frame through a Lorentz boost with velocity
$\tilde\beta=(x_+-x_-)/(x_++x_-)$, and
\begin{equation}
x(x_+,x_-,x^\prime,\cos\theta^\prime)=\tilde\gamma\left(\sqrt{\tau}x^\prime
+\tilde\beta\sqrt{\tau x^{\prime2}-\rho_H}\cos\theta^\prime\right),
\end{equation}
with $\tilde\gamma=1/\sqrt{1-{\tilde\beta}^2}$.\footnote{%
We denote the relativistic boost velocity and its $\gamma$ factor by
$\tilde\beta$ and $\tilde\gamma$, respectively, because $\beta$ and $\gamma$
are reserved for other quantities in this paper.}
Integrating Eq.~(\ref{eq:master}) over $\cos\theta^\prime$, we have
\begin{eqnarray}
\frac{\diff\sigma_\mathrm{ISR}}{\diff x}(x,s)
&=&\int\limits_x^1\diff x^\prime
\left(\int\limits_{x/x^\prime}^1\diff x_+
\int\limits_{a_-/(x^\prime-a_+/x_+)}^{a_+/(x^\prime-a_-/x_+)}\diff x_-
+\int\limits_{x/x^\prime}^1\diff x_-
\int\limits_{a_-/(x^\prime-a_+/x_-)}^{a_+/(x^\prime-a_-/x_-)}\diff x_+\right)
\label{eq:isr}\\
&&\times\frac{2D_{e^+}(x_+,s)D_{e^-}(x_-,s)}
{|x_+-x_-|\sqrt{x^{\prime2}-\rho_H/\tau}}\,
\frac{\diff^2\sigma}{\diff x^\prime\,\diff\cos\theta^\prime}
\left(x^\prime,
\frac{2x-(x_++x_-)x^\prime}{(x_+-x_-)\sqrt{x^{\prime2}-\rho_H/\tau}},
\tau s\right),
\nonumber
\end{eqnarray}
where
\begin{equation}
a_\pm=\frac{x\pm\sqrt{x^2-\rho_H}}{2}.
\label{eq:xpm}
\end{equation}
Notice that the second pair of integrations enclosed within the parentheses in
Eq.~(\ref{eq:isr}) merely duplicates the first one if
$D_{e^+}(x,s)=D_{e^-}(x,s)$.

Detailed numerical inspection of our specific application reveals that, to
very good approximation, we may substitute in Eq.~(\ref{eq:isr})
\begin{equation}
D_{e^+}(x_+,s)D_{e^-}(x_-,s)\approx\delta(1-x_+)H_{e^+e^-}(x_-,s),
\label{eq:del}
\end{equation}
which trivially satisfies Eq.~(\ref{eq:h}), so as to eliminate the second term
within the parentheses in Eq.~(\ref{eq:isr}) and to save one integration in
the first one.
This implies that the whole ISR is emitted by the electron alone, while the
positron stays idle, or {\it vice versa}.
A further simplification may be obtained by casting Eq.~(\ref{eq:isr}) into a
form that contains on the r.h.s.\ $(\diff\sigma/\diff x^\prime)(x^\prime,s)$,
{\it i.e.}\ the cross section at the $e^+e^-$ c.m.\ energy integrated over the
polar angle.
To this end, we use the approximations
\begin{eqnarray}
\frac{\diff^2\sigma}{\diff x^\prime\,\diff\cos\theta^\prime}
(x^\prime,\cos\theta^\prime,\tau s)&\approx&
\frac{3}{8}(1+\cos^2\theta^\prime)
\frac{\diff\sigma}{\diff x^\prime}(x^\prime,\tau s)
\nonumber\\
&\approx&\frac{3}{8}(1+\cos^2\theta^\prime)
\frac{\sigma(\tau s)}{\sigma(s)}\,
\frac{\diff\sigma}{\diff x^\prime}(x^\prime,s).
\label{eq:app}
\end{eqnarray}
The approximation in the first line of Eq.~(\ref{eq:app}) may be justified by
observing that, in Eq.~(\ref{eq:xstheta}), $\diff\sigma^L/\diff x$ is
suppressed for $m^2\ll s$, being zero at LO in the ZM approach, and that the
contribution from $\diff\sigma^A/\diff x$ vanishes upon integration over
$\cos\theta^\prime$.
The approximation in the second line of Eq.~(\ref{eq:app}) faithfully
describes the leading power-like dependence on the c.m.\ energy, but
disregards the scaling violations of the FFs, which are just logarithmic.
For simplicity, we evaluate the factor $\sigma(\tau s)/\sigma(s)$ in
Eq.~(\ref{eq:app}) at LO in the GM approach neglecting the contribution from
$Z$-boson exchange, as
\begin{equation}
\frac{\sigma(\tau s)}{\sigma(s)}=\frac{1}{\tau}\,
\frac{1+\rho/(2\tau)}{1+\rho/2}\sqrt{\frac{1-\rho/\tau}{1-\rho}}.
\label{eq:rat}
\end{equation}
On the one hand, the NLO corrections largely cancel out in this cross section
ratio; on the other hand, the $Z$-boson contribution is suppressed for
$\sqrt s\ll m_Z$, as explained in Appendix~\ref{sec:appa}.
Substituting Eqs.~(\ref{eq:del}) and (\ref{eq:app}) into Eq.~(\ref{eq:isr}),
we obtain our working formula
\begin{equation}
\frac{\diff\sigma_\mathrm{ISR}}{\diff x}(x,s)
\approx\int\limits_{x}^1\diff x^\prime
\frac{\diff\sigma}{\diff x^\prime}(x^\prime,s)
\int\limits_{a_-/(x^\prime-a_+)}^{a_+/(x^\prime-a_-)}\diff\tau
\frac{2H_{e^+e^-}(\tau,s)}{(1-\tau)\sqrt{x^{\prime2}-\rho_H/\tau}}\,
\frac{3}{8}(1+\cos^2\theta^\prime)\frac{\sigma(\tau s)}{\sigma(s)},
\label{eq:final}
\end{equation}
where
\begin{equation}
\cos\theta^\prime=\frac{2x-(1+\tau)x^\prime}
{(1-\tau)\sqrt{x^{\prime2}-\rho_H/\tau}}
\end{equation}
and $\sigma(\tau s)/\sigma(s)$ is given in Eq.~(\ref{eq:rat}).
In Eq.~(\ref{eq:final}), we insert the NLO expression for
$(\diff\sigma/\diff x^\prime)(x^\prime,s)$.
Strictly speaking, we should then also evaluate Eq.~(\ref{eq:rat}) at NLO.
However, the omitted correction is insignificant, as explained below
Eq.~(\ref{eq:rat}).
We employ this formalism also in the ZM approach, except that we then set
$\rho=0$ in Eq.~(\ref{eq:rat}) for consistency. 

\begin{figure}[ht]
\begin{center}
\epsfig{file=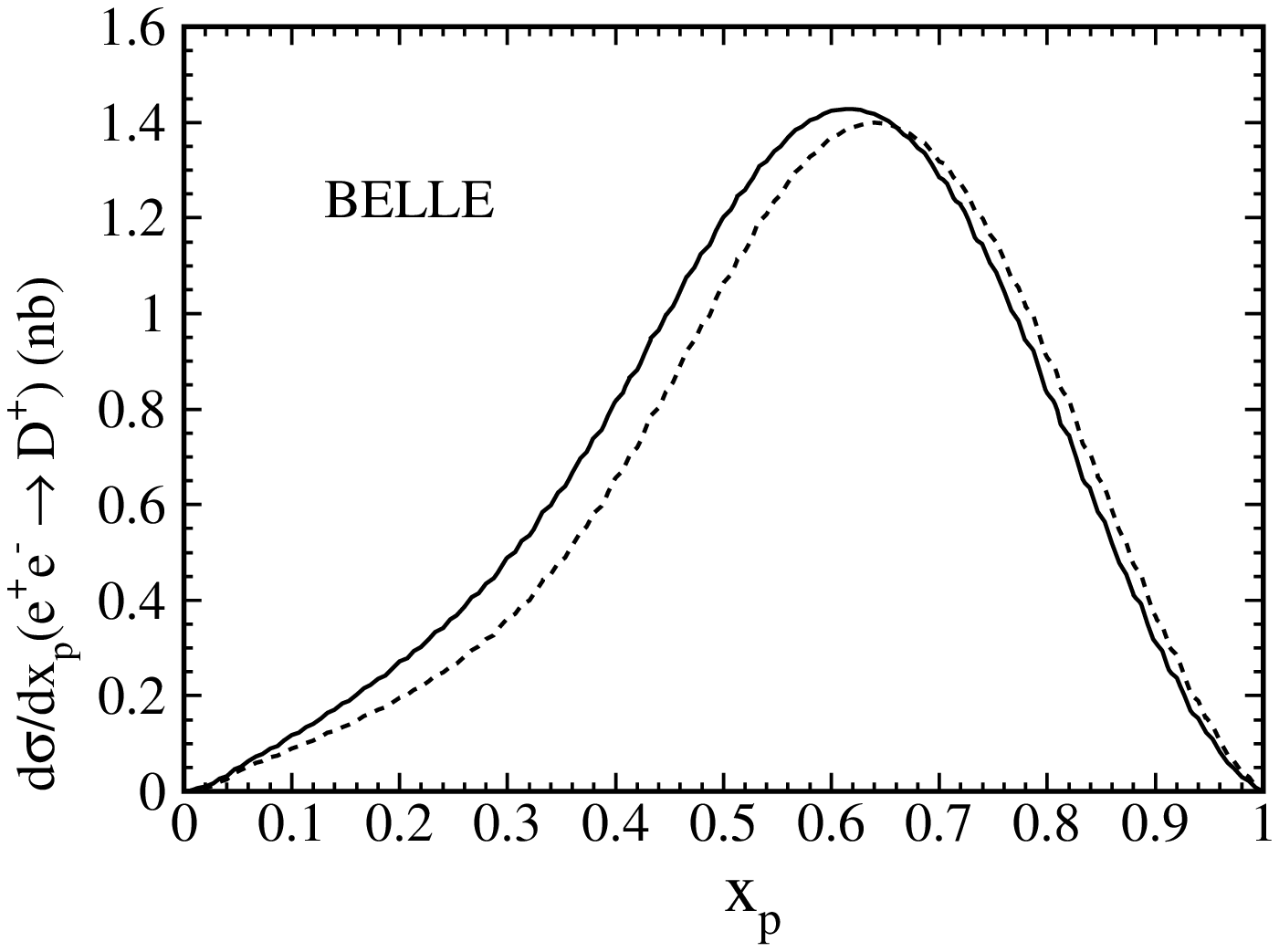,width=0.90\textwidth}
\end{center}
\caption{\label{fig:ISRPlot}$x_p$ distributions of $e^+e^-\to D^++X$ at
$\sqrt s=10.52$~GeV evaluated in the GM approach with the FFs from the joint
fit to the Belle \cite{Seuster:2005tr} and CLEO \cite{Artuso:2004pj} data
including ISR corrections (solid line) and corresponding result with the
latter subtracted (dashed line).}
\end{figure}
The effect of ISR on the $x_p$ distribution of $e^+e^- \rightarrow D^++X$
measured by Belle \cite{Seuster:2005tr} and CLEO \cite{Artuso:2004pj} is
studied for the GM approach in Fig.~\ref{fig:ISRPlot}, where the result of the
ISR-corrected fit to these data is compared with the corresponding result
where the ISR corrections are subtracted.
As already observed in Ref.~\cite{Cacciari:2005uk}, we find that the spectrum
is shifted to larger values of $x_p$, {\it i.e.}\ it becomes harder and lower
at the peak when the ISR corrections are subtracted.
This is expected, since ISR reduces the available hadronic c.m.\ energy, which
softens the spectrum and increases the cross section.

At LEP1 energy, the shift of the spectrum due to ISR is negligible.
This may be understood by observing that ISR shifts the c.m.\ energy available
for the hard scattering to values below the $Z$-boson resonance, where the
cross section is greatly reduced.
In the analysis described below, we thus only include ISR corrections in our
theoretical description of the Belle \cite{Seuster:2005tr} and CLEO
\cite{Artuso:2004pj} data, while we neglect them in connection with the ALEPH
\cite{Barate:1999bg} and OPAL \cite{Alexander:1996wy,Ackerstaff:1997ki} data.

\section{Results}
\label{sec:three}

As experimental input for our fits, we use the $x_p$ distributions of $D^0$,
$D^+$, and $D^{*+}$ production in the continuum at $\sqrt s=10.52$~GeV for
$0.08<x_p<0.94$ from Belle \cite{Seuster:2005tr} and for $0.20<x_p<0.95$ from
CLEO \cite{Artuso:2004pj}, which we correct for ISR as described in
Section~\ref{sec:twoc}, and the $x$ distributions of $D^0$, $D^+$
\cite{Alexander:1996wy}, and $D^{*+}$ \cite{Ackerstaff:1997ki,Barate:1999bg}
production on the $Z$-boson resonance at $\sqrt s=91.2$~GeV from ALEPH and
OPAL.
We received the Belle data in numerical form via private communication
\cite{Seuster:pc}. 
Belle \cite{Seuster:2005tr} also provide data from the $\Upsilon(5S)$
resonance outside the $B$-meson decay region, which we leave aside.

In Refs.~\cite{Alexander:1996wy,Ackerstaff:1997ki,Barate:1999bg}, the absolute
cross section distributions in $x$ are normalized to the total hadronic cross
section and include the branching fractions of the decays used to identify
the $H_c$ mesons, namely $D^0\to K^-\pi^+$, $D^+\to K^-\pi^+\pi^+$, and
$D^{*+}\to D^0\pi^+$ followed by $D^0\to K^-\pi^+$, respectively.
Therefore, we multiply our predictions by $1/\sigma_\mathrm{tot}$, where, for
simplicity, we evaluate $\sigma_\mathrm{tot}$ from Eq.~(\ref{eq:sigtot})
exploiting the insignificance of quark mass effects at LEP1 energy, and divide
the experimental data by the respective decay branching fractions.
For consistency, we adopt the very values of the latter that are used in
Refs.~\cite{Alexander:1996wy,Ackerstaff:1997ki,Barate:1999bg}.
These read $B(D^0\to K^-\pi^+)=(3.84\pm0.13)\%$ and
$B(D^+\to K^-\pi^+\pi^+)=(9.1\pm0.6)\%$ \cite{Montanet:1994xu} for
Ref.~\cite{Alexander:1996wy},
$B(D^{*+}\to D^0\pi^+)=(68.3\pm1.4)\%$ and
$B(D^0\to K^-\pi^+)=(3.83\pm0.12)\%$ \cite{Barnett:1996hr} for
Ref.~\cite{Ackerstaff:1997ki}, and
$B(D^{*+}\to D^0\pi^+)=(68.3\pm1.4)\%$ and
$B(D^0\to K^-\pi^+)=(3.85\pm0.09)\%$ \cite{Caso:1998tx} for
Ref.~\cite{Barate:1999bg}, respectively.
The $b$-tagged samples of Refs.~\cite{Alexander:1996wy,Ackerstaff:1997ki} are
treated in the same way as the full cross sections, while the one in
Ref.~\cite{Barate:1999bg} needs to be multiplied by $R_bf(b\to D^{*\pm})$,
where $R_b=\Gamma(Z\to b\overline{b})/\Gamma(Z\to\mathrm{hadrons})$ is the
fraction of $b$-tagged events in the full hadronic sample and
$f(b\to D^{*\pm})$ is the probability of a $b$ quark to hadronise into a
$D^{*\pm}$ meson.
We adopt the value $R_bf(b\to D^{*+})=(4.66\pm0.51)\%$ determined in
Ref.~\cite{Barate:1999bg}.

We take $\alpha$ in Eq.~(\ref{eq:sig0}) to be the running fine-structure
constant, which is particularly important because it appears there in squared
form.
At $\sqrt s=10.52$~GeV, we have $1/\alpha\approx132$ \cite{Hagiwara:1994pw}.
Of course, this effect cancels out in the normalized cross sections considered
in Refs.~\cite{Alexander:1996wy,Ackerstaff:1997ki,Barate:1999bg}.

As already mentioned in Section~\ref{sec:one}, we perform for each hadron
species a combined fit to the Belle, CLEO, ALEPH, and OPAL data (global fit),
a separate fit to the $B$-factory data (Belle/CLEO), and a separate fit to the
$Z$-factory data (ALEPH/OPAL).
We perform fits both for $m\neq0$ (GM) and for $m=0$ (ZM).

We parameterize the $z$ distributions of the $c$ and $b$ quark FFs at their
starting scales $\mu_0$ as suggested by Bowler \cite{Bowler:1981sb}, as
\begin{equation}
D_Q^{H_c}(z,\mu_0) = N z^{-(1+ \gamma^2)} (1-z)^a {\rm e}^{-\gamma^2 /z },
\label{eq:Bowler}
\end{equation}
with three free parameters, $N$, $a$, and $\gamma$.
This parameterization yielded the best fit to the Belle data
\cite{Seuster:2005tr} in a comparative analysis using the Monte-Carlo event
generator JETSET/PYTHIA.

Specifically, our fitting procedure is as follows.
At the scale $\mu_f=m_c=1.5$~GeV, the $c$-quark FF is taken to be of the form
specified in Eq.~(\ref{eq:Bowler}), while the FFs of the light quarks $q$
($q = u, d, s$) and the gluon are set to zero.
Then these FFs are evolved to higher scales using the DGLAP equations in
Eq.~(\ref{eq:DGLAP}) at NLO with $n_f=4$ active quark flavors and an
appropriate value $\Lambda_{\overline{\rm MS}}^{(4)}$ of the asymptotic scale
parameter. 
When the scale reaches the threshold value $\mu_f=m_b=5.0$~GeV, the bottom
flavor is activated and its FF is introduced in the Bowler form of
Eq.~(\ref{eq:Bowler}).
The evolution to higher scales is then performed with $n_f=5$ and the value
$\Lambda_{\overline{\rm MS}}^{(5)}$ is properly matched to
$\Lambda_{\overline{\rm MS}}^{(4)}$.
Including $\Lambda_{\overline{\rm MS}}^{(4)}$ among the fit parameters, it
turns out to be feebly constrained by the fit.
Therefore, we adopt the value $\Lambda_{\overline{\rm MS}}^{(5)}=221$~MeV from
Ref.~\cite{Yao:2006px} and adjust the value of
$\Lambda_{\overline{\rm MS}}^{(4)}$ accordingly, to be
$\Lambda_{\overline{\rm MS}}^{(4)}=321$~MeV.

\begin{figure}[ht]
\begin{center}
\begin{tabular}{ll}
\parbox{0.45\textwidth}{
\epsfig{file=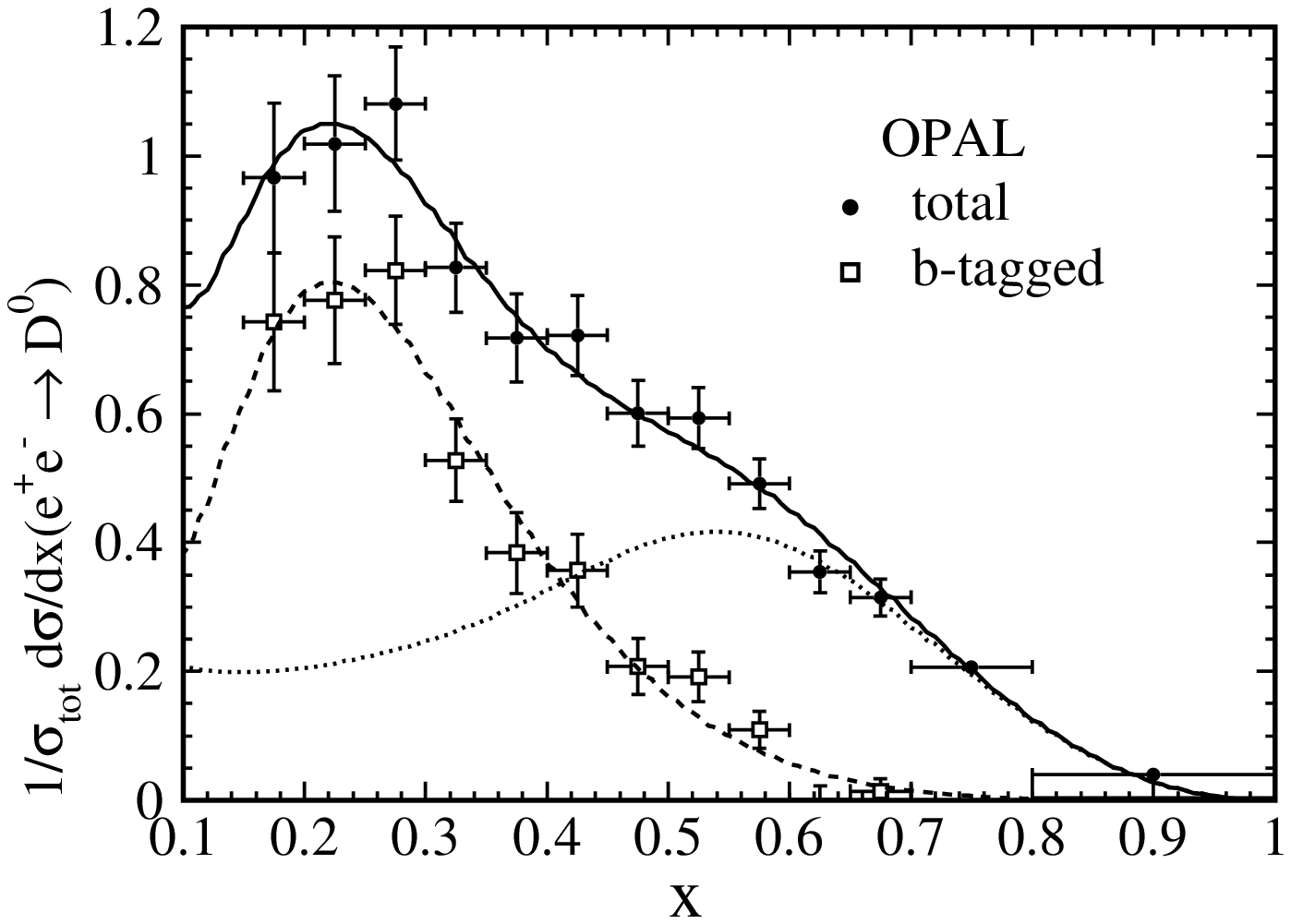,width=0.45\textwidth}
}
&
\parbox{0.45\textwidth}{
\epsfig{file=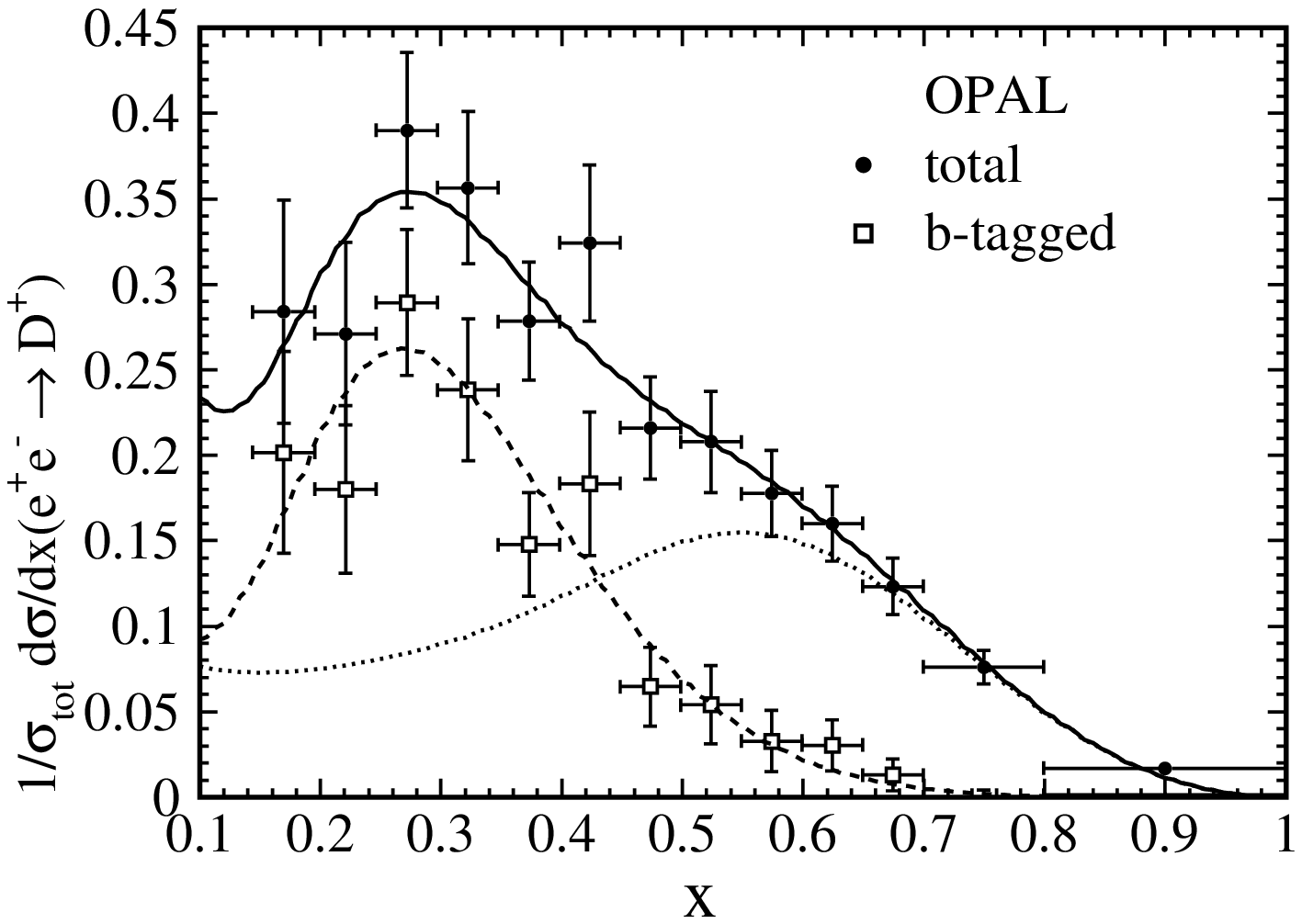,width=0.45\textwidth}
}
\\
(a) & (b)\\
\parbox{0.45\textwidth}{
\epsfig{file=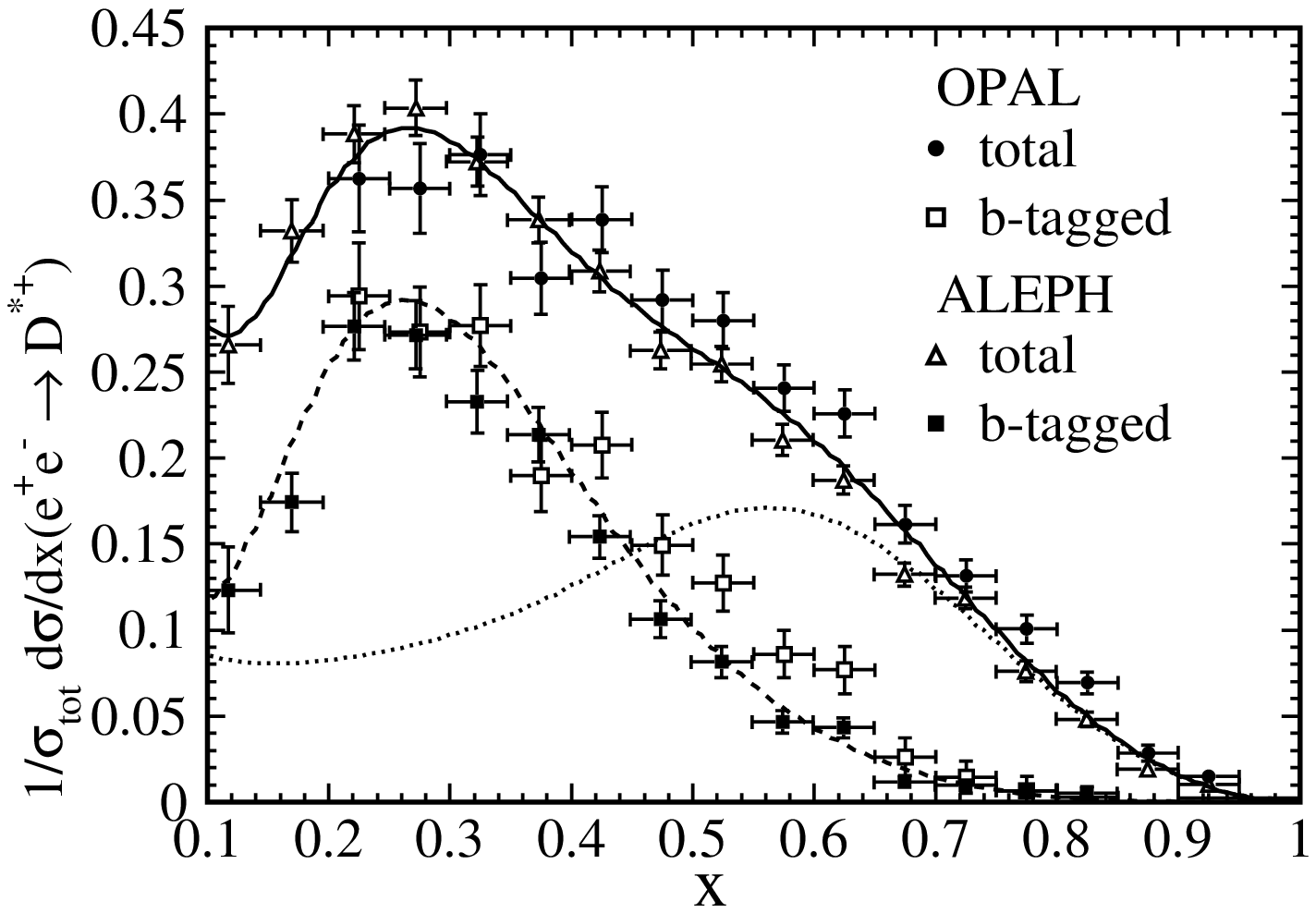,width=0.45\textwidth}
}
&
\\
(c) &
\end{tabular}
\end{center}
\caption{\label{Fig:OPALFits}Normalized $x$ distributions of (a) $D^0$, (b)
$D^+$, and (c) $D^{*+}$ mesons from OPAL
\cite{Alexander:1996wy,Ackerstaff:1997ki} and ALEPH \cite{Barate:1999bg}
compared to the respective fits in the GM approach from
Tables~\ref{Tab:D0Fits}--\ref{Tab:D+stFits}.
The dotted, dashed, and solid lines refer to the $c$-quark-initiated,
$b$-quark-initiated, and total contributions, respectively.}
\end{figure}

\begin{table}
\caption{\label{Tab:D0Fits}Values of fit parameters for the $D^0$ meson
resulting from the Belle/CLEO, OPAL, and global fits in the GM approach
together with the values of $\overline{\chi^2}$ achieved.}
\begin{center}
\begin{tabular}{|l|c|c|c|c|} \hline
 & Belle/CLEO-GM & OPAL-GM & global-GM \\ 
\hline
$N_c$ & $1.51 \cdot 10^7$ & $4.42 \cdot 10^4$ & $8.80 \cdot 10^6$ \\
$a_c$ & 1.56 & 1.52 & 1.54 \\
$\gamma_c$ & 3.64 & 2.83 & 3.58 \\
\hline
$N_b$ & 13.5 & 13.5 & 78.5 \\
$a_b$ & 3.98 & 3.98 & 5.76 \\
$\gamma_b$ & 0.921 & 0.921 & 1.14 \\
\hline
$\overline{\chi ^2}$ & 3.15 & 0.794 & 4.03 \\
\hline
\end{tabular}
\end{center}
\end{table}
\begin{table}
\caption{\label{Tab:D+Fits}Values of fit parameters for the $D^+$ meson
resulting from the Belle/CLEO, OPAL, and global fits in the GM approach
together with the values of $\overline{\chi^2}$ achieved.}
\begin{center}
\begin{tabular}{|l|c|c|c|c|} \hline
 & Belle/CLEO-GM & OPAL-GM & global-GM \\ 
\hline
$N_c$ & $5.66 \cdot 10^5$ & $2.82 \cdot 10^4$ & $5.67 \cdot 10^5$ \\
$a_c$ & 1.15 & 1.49 & 1.16 \\
$\gamma_c$ & 3.39 & 2.92 & 3.39 \\
\hline
$N_b$ & 18.8 & 18.8 & 185 \\
$a_b$ & 4.71 & 4.71 & 7.08 \\
$\gamma_b$ & 1.17 & 1.17 & 1.42 \\
\hline
$\overline{\chi ^2}$ & 1.30 & 0.509 & 1.99 \\
\hline
\end{tabular}
\end{center}
\end{table}
\begin{table}
\caption{\label{Tab:D+stFits}Values of fit parameters for the $D^{*+}$ meson
resulting from the Belle/CLEO, ALEPH/OPAL, and global fits in the GM
approach together with the values of $\overline{\chi^2}$ achieved.}
\begin{center}
\begin{tabular}{|l|c|c|c|c|c|} \hline
 & Belle/CLEO-GM & ALEPH/OPAL-GM & global-GM \\ 
\hline
$N_c$ & $1.33 \cdot 10^7$ & $4.58 \cdot 10^4$ & $1.10 \cdot 10^7$ \\
$a_c$ & 0.992 & 1.38 & 1.07 \\
$\gamma_c$ & 3.84 & 3.00 & 3.81 \\
\hline
$N_b$ & 6.67 & 6.67 & 14.0 \\
$a_b$ & 3.28 & 3.28 & 3.85 \\
$\gamma_b$ & 1.04 & 1.04 & 1.14 \\
\hline
$\overline{\chi ^2}$ & 3.74 & 2.06 & 6.90 \\
\hline
\end{tabular}
\end{center}
\end{table}
We first describe the Belle/CLEO fits in the GM approach.
In the Belle and CLEO data, all charmed hadrons coming from $B$-meson decays
are excluded, so that there is no need to include $b\to H_c$ fragmentation.
On the other hand, the ALEPH and OPAL data each come as two sets:
the sample of $H_c$ hadrons produced by the decays of $b$ hadrons from
$Z\to b\bar{b}$ ($b$ tagged) and the total sample of $H_c$ hadrons, also
including those from direct production in $Z\to c\bar{c}$ and from light-quark
and gluon fragmentation (total).
Since we wish to test the Belle/CLEO FFs fits through comparisons to LEP1
data, we must include the $b\to H_c$ transitions in an appropriate way.
For this purpose, we first fit the $D^0$ and $D^+$ data from OPAL and the
$D^{*+}$ data from ALEPH and OPAL.
The resulting values of the fit parameters and of $\chi^2$ per degree of
freedom, $\overline{\chi^2}$, for the $D^0$, $D^+$, and $D^{*+}$ mesons are
given in Tables~\ref{Tab:D0Fits}--\ref{Tab:D+stFits}, respectively. 
The goodness of these fits may also be judged from
Figs.~\ref{Fig:OPALFits}(a)--(c), respectively.
We observe from Fig.~\ref{Fig:OPALFits}(c) that the $D^{*+}$ data from ALEPH
and OPAL are only moderately compatible, as was already noticed in
Ref.~\cite{Binnewies:1997gz}.
This explains why the corresponding fit has a larger value of
$\overline{\chi^2}$ than those for the $D^0$ and $D^+$ mesons.

\begin{figure}[ht]
\begin{center}
\begin{tabular}{ll}
\parbox{0.45\textwidth}{
\epsfig{file=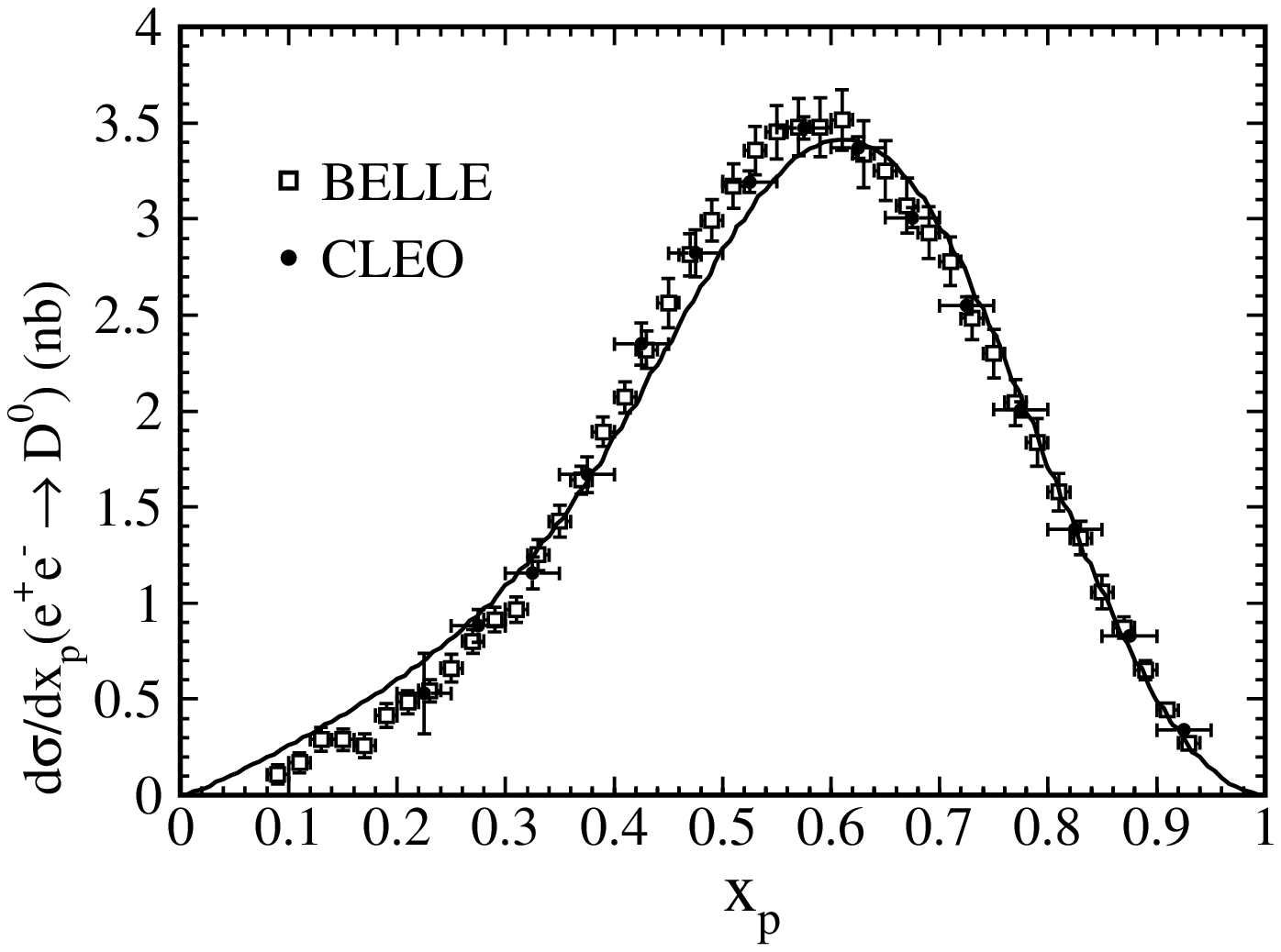,width=0.45\textwidth}
}
&
\parbox{0.45\textwidth}{
\epsfig{file=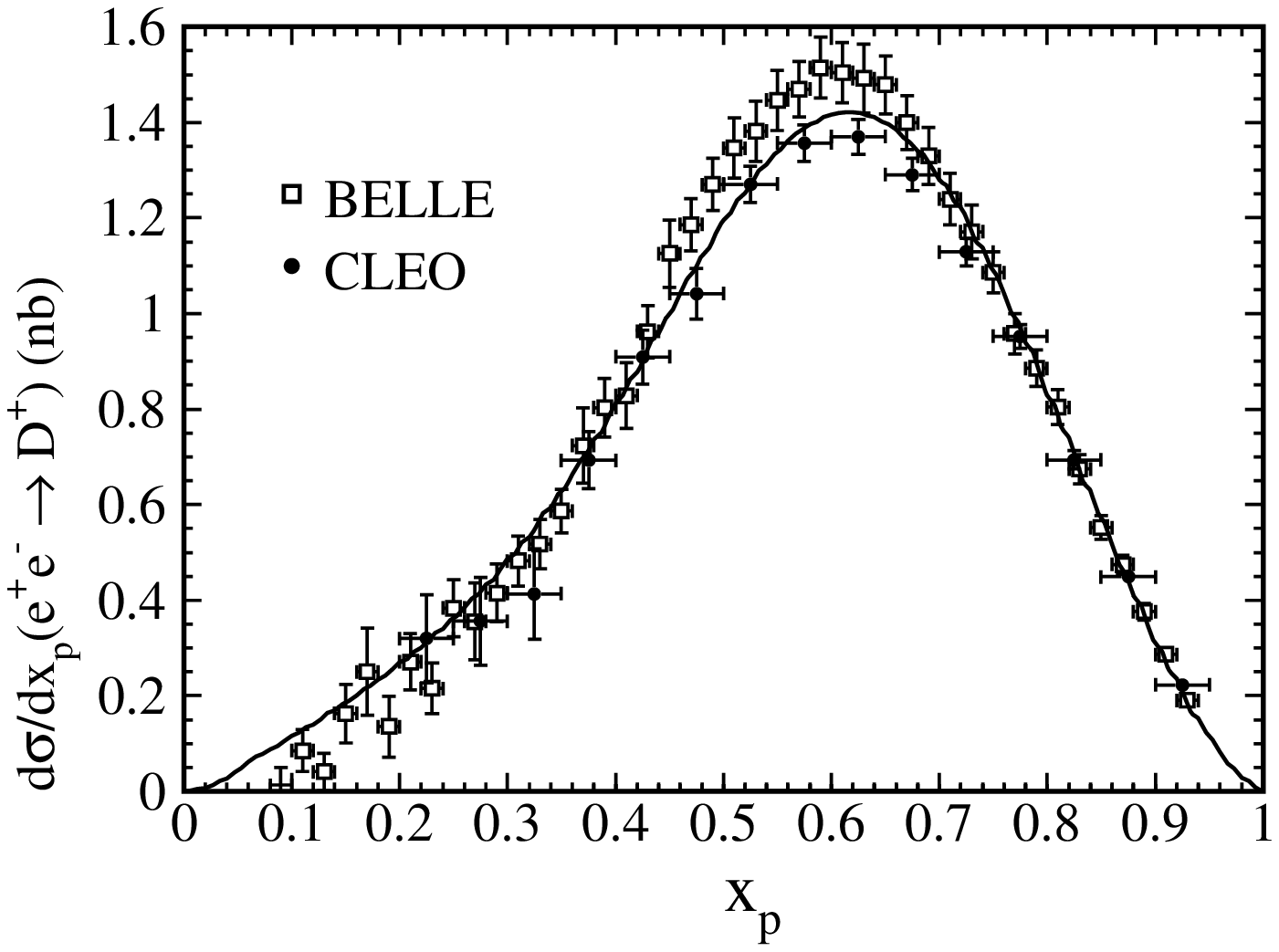,width=0.45\textwidth}
}
\\
(a) & (b)\\
\parbox{0.45\textwidth}{
\epsfig{file=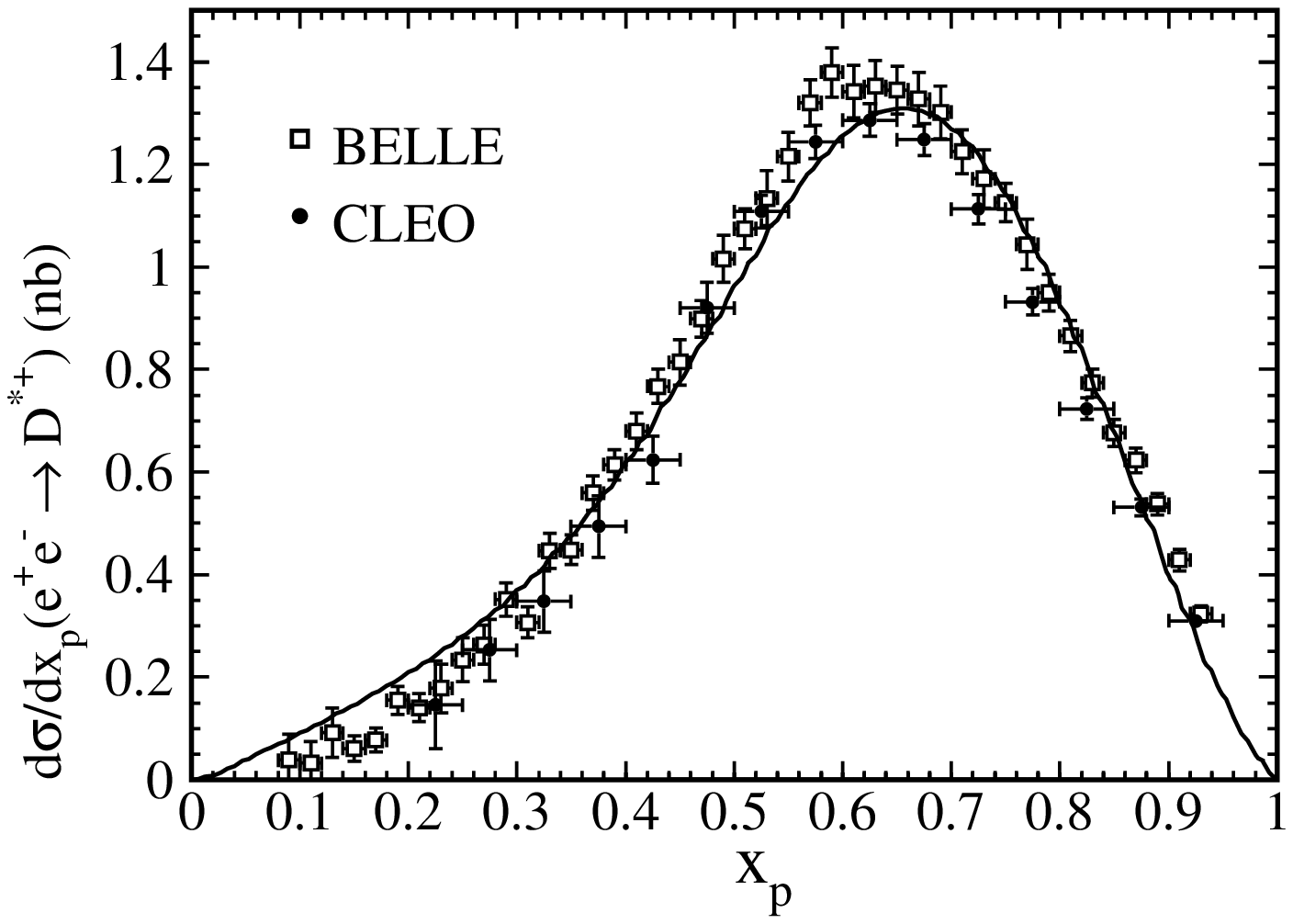,width=0.45\textwidth}
}
&
\\
(c) & 
\end{tabular}
\end{center}
\caption{\label{Fig:BELLEFits}$x_p$ distributions of (a) $D^0$, (b) $D^+$, and
(c) $D^{*+}$ mesons from Belle \cite{Seuster:2005tr} and CLEO
\cite{Artuso:2004pj} compared to the respective fits in the GM approach from
Tables~\ref{Tab:D0Fits}--\ref{Tab:D+stFits}.}
\end{figure}

In a second step, we use the values of $N_b$, $a_b$, and $\gamma_b$ thus
obtained as rigid input for the fits to the Belle and CLEO data.
While the $b\to H_c$ transitions are excluded from the final states as
explained above, their FFs still influence the fits through the DGLAP
evolution from $\mu_f=m_b$ to $\mu_f=\sqrt s$.
These fits yield new values for $N_c$, $a_c$, and $\gamma_c$, which are also
included in Tables~\ref{Tab:D0Fits}--\ref{Tab:D+stFits} together with the
values of $\overline{\chi^2}$ achieved.
The $\overline{\chi^2}$ values of the fits discussed so far are all
acceptable, except perhaps for the fits to the $D^0$ and $D^{*+}$ data from
Belle and CLEO, which yield $\overline{\chi^2}$ values in excess of 3.
In Figs.~\ref{Fig:BELLEFits}(a)--(c), the $D^0$, $D^+$, and $D^{*+}$ data from
Belle and CLEO are compared with the respective fit results in order to assess
the goodness of the latter.
We observe from Figs.~\ref{Fig:BELLEFits}(a)--(c) that the agreement is best
in the $D^+$ case, which is also reflected in
Tables~\ref{Tab:D0Fits}--\ref{Tab:D+stFits}.

\begin{figure}[ht]
\begin{center}
\begin{tabular}{ll}
\parbox{0.45\textwidth}{
\epsfig{file=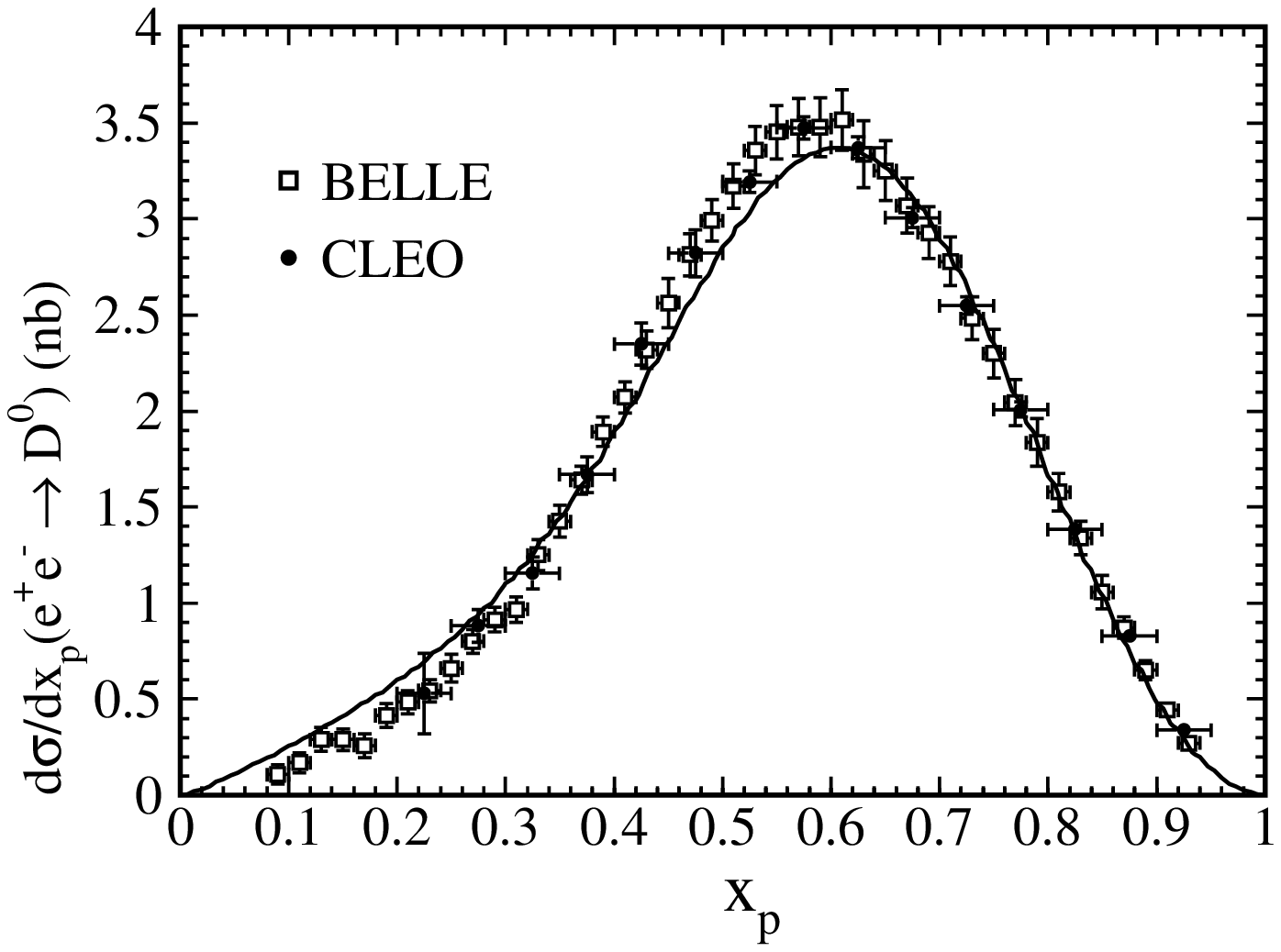,width=0.45\textwidth}
}
&
\parbox{0.45\textwidth}{
\epsfig{file=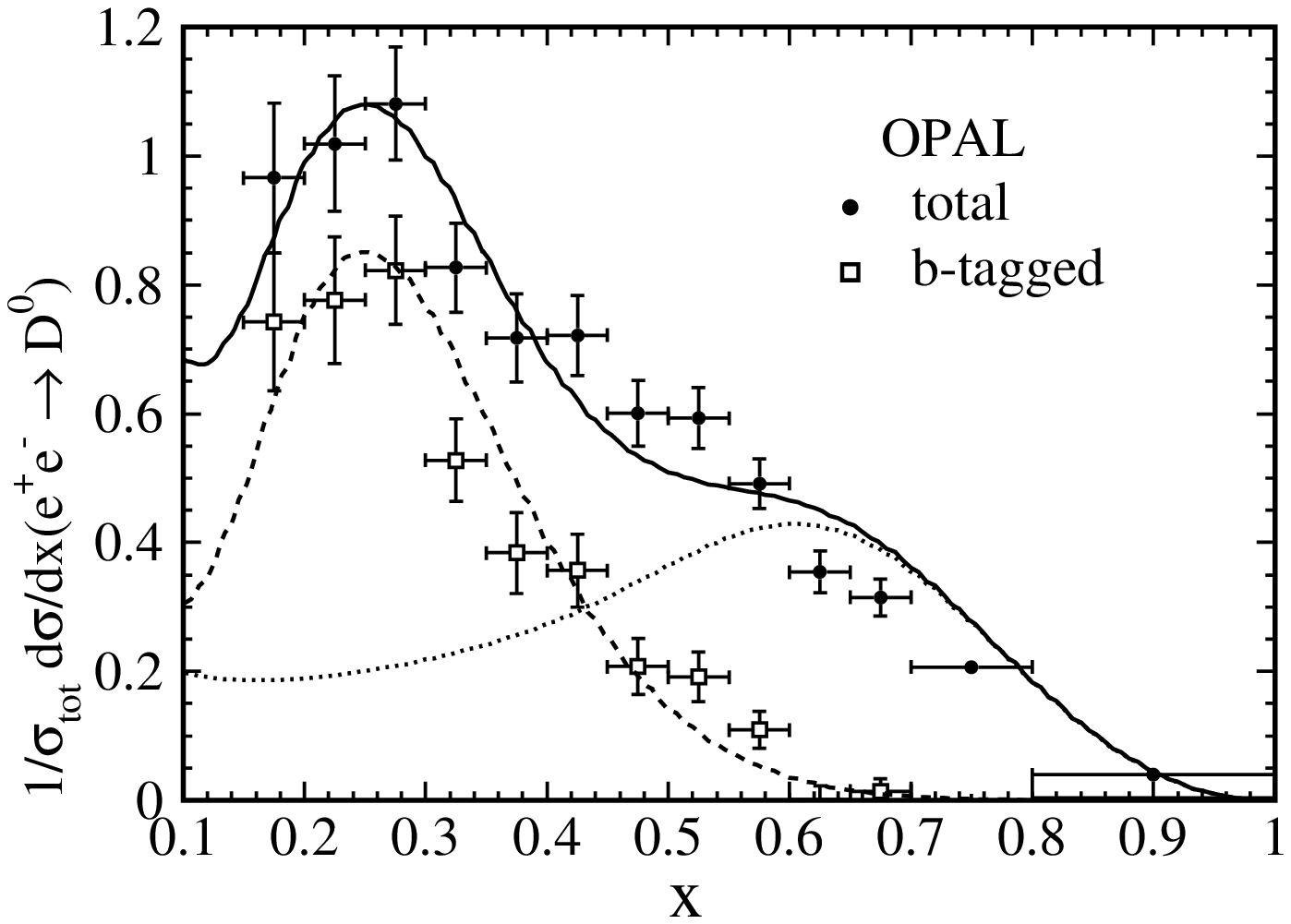,width=0.45\textwidth}
}
\\
(a) & (b)
\end{tabular}
\end{center}
\caption{\label{Fig:D0Global}$x_p$ distributions of $D^0$ mesons from (a)
Belle \cite{Seuster:2005tr} and CLEO \cite{Artuso:2004pj}, and (b) normalized
$x$ distribution of $D^0$ mesons from OPAL \cite{Alexander:1996wy} compared to
the global fit from Table~\ref{Tab:D0Fits}.
In frame~(b), the dotted, dashed, and solid lines refer to the
$c$-quark-initiated, $b$-quark-initiated, and total contributions,
respectively.}
\end{figure}

\begin{figure}[ht]
\begin{center}
\begin{tabular}{ll}
\parbox{0.45\textwidth}{
\epsfig{file=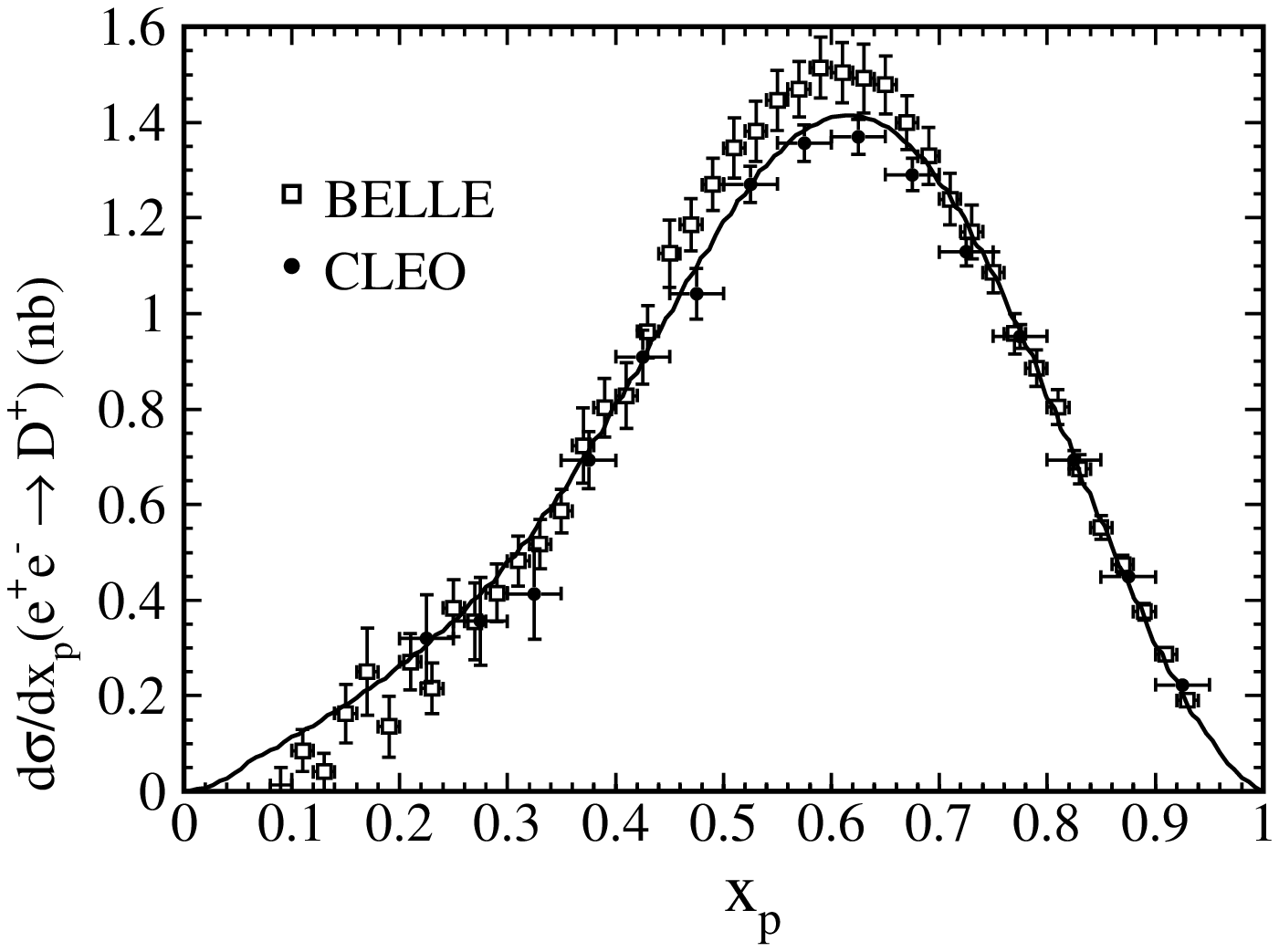,width=0.45\textwidth}
}
&
\parbox{0.45\textwidth}{
\epsfig{file=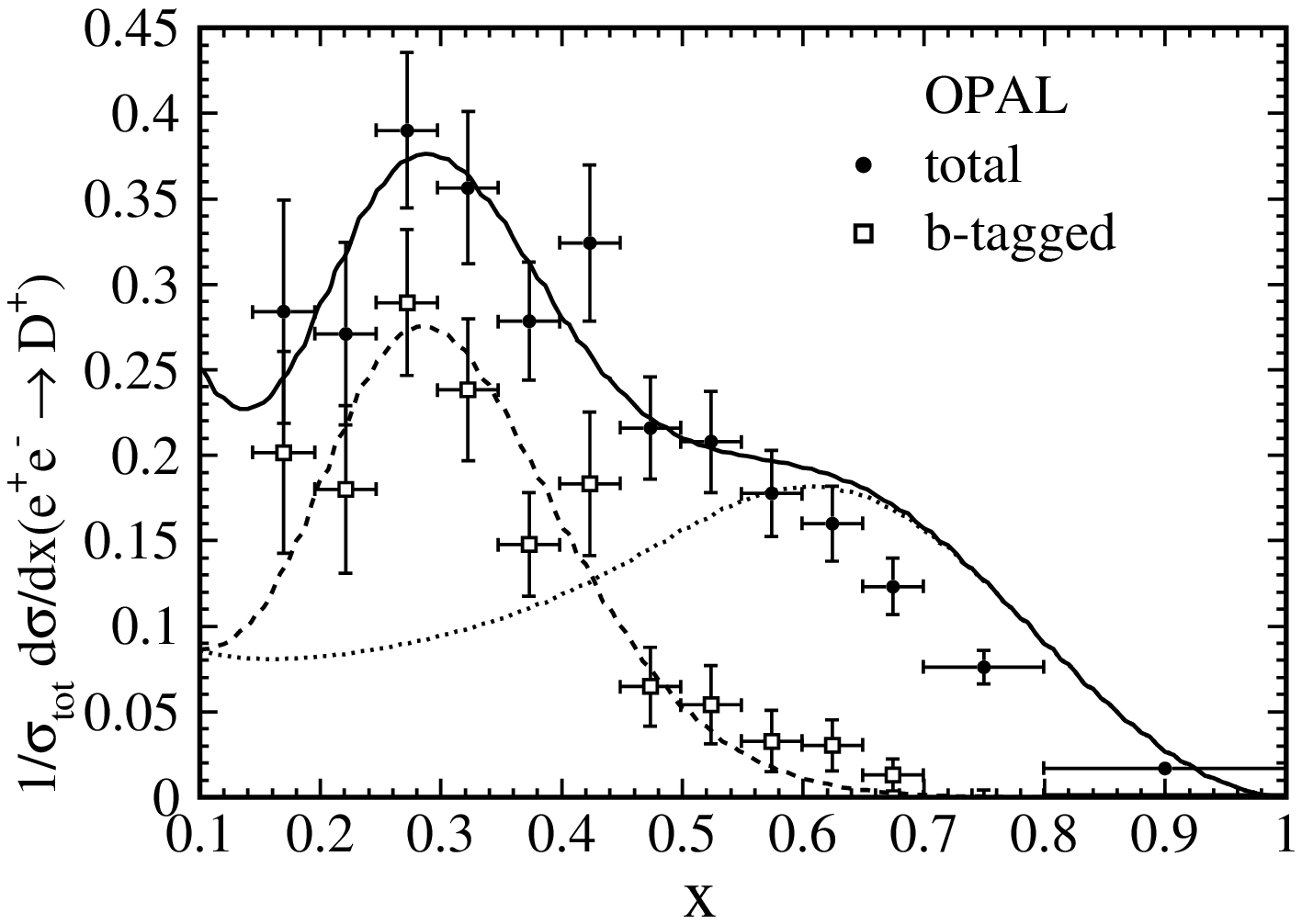,width=0.45\textwidth}
}
\\
(a) & (b)
\end{tabular}
\end{center}
\caption{\label{Fig:D+Global}$x_p$ distributions of $D^+$ mesons from (a)
Belle \cite{Seuster:2005tr} and CLEO \cite{Artuso:2004pj}, and (b) normalized
$x$ distribution of $D^+$ mesons from OPAL \cite{Alexander:1996wy} compared to
the global fit from Table~\ref{Tab:D+Fits}.
In frame~(b), the dotted, dashed, and solid lines refer to the
$c$-quark-initiated, $b$-quark-initiated, and total contributions,
respectively.}
\end{figure}

\begin{figure}[ht]
\begin{center}
\begin{tabular}{ll}
\parbox{0.45\textwidth}{
\epsfig{file=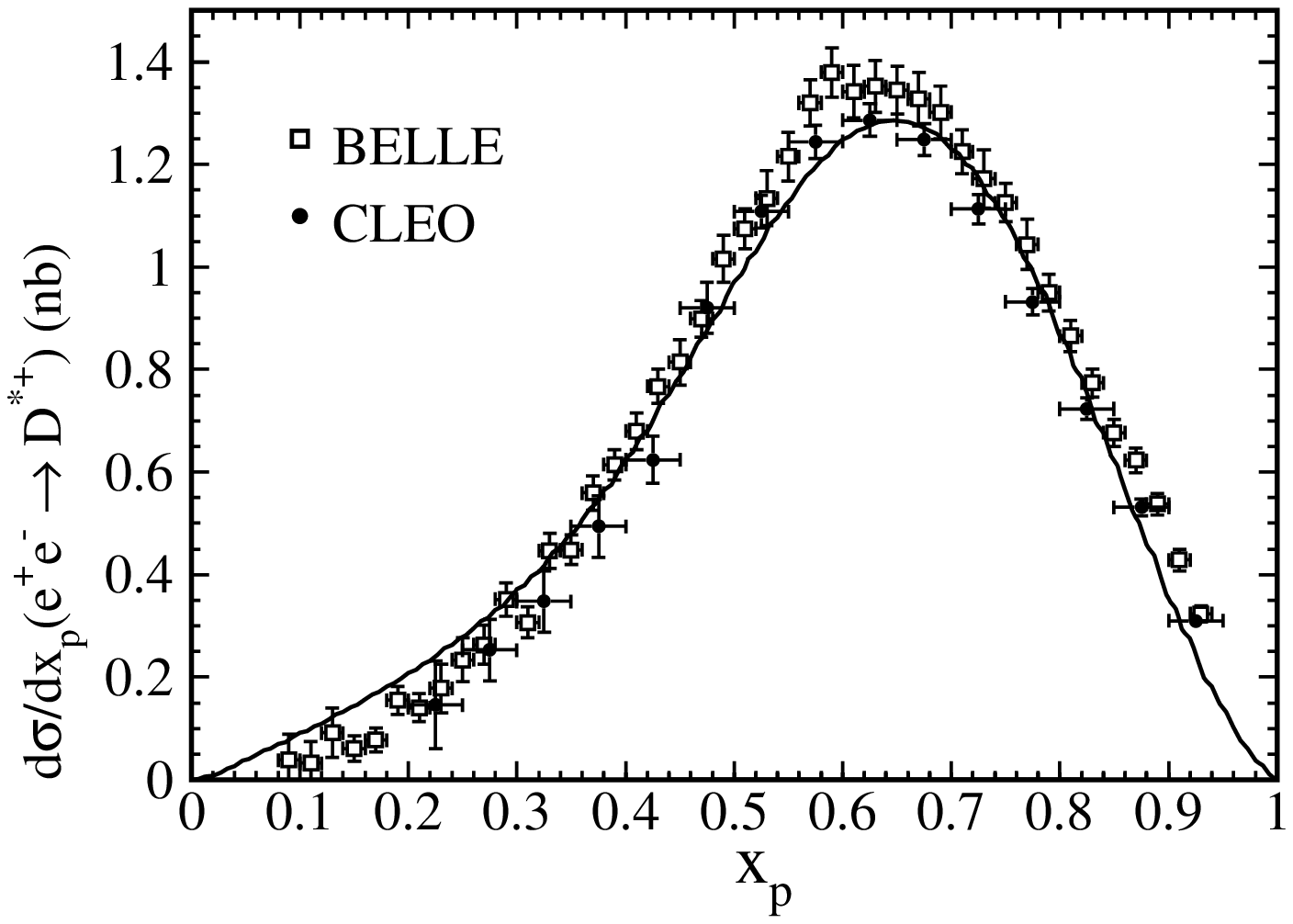,width=0.45\textwidth}
}
&
\parbox{0.45\textwidth}{
\epsfig{file=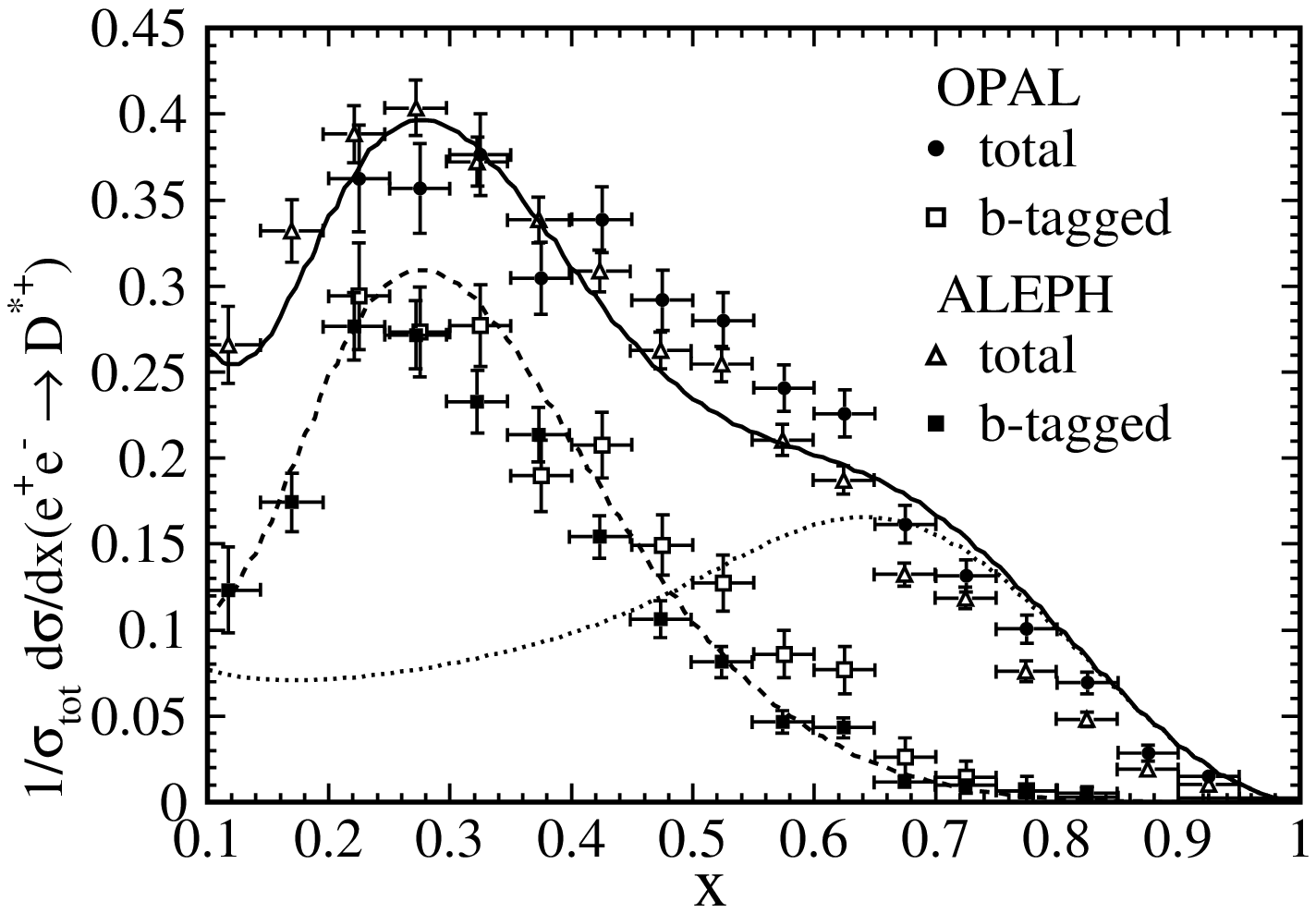,width=0.45\textwidth}
}
\\
(a) & (b)
\end{tabular}
\end{center}
\caption{\label{Fig:D+stGlobal}$x_p$ distributions of $D^{*+}$ mesons from (a)
Belle \cite{Seuster:2005tr} and CLEO \cite{Artuso:2004pj}, and (b) normalized 
$x$ distributions of $D^{*+}$ mesons from ALEPH \cite{Barate:1999bg}, and OPAL
\cite{Ackerstaff:1997ki} compared to the global fit from
Table~\ref{Tab:D+stFits}.
In frame~(b), the dotted, dashed, and solid lines refer to the
$c$-quark-initiated, $b$-quark-initiated, and total contributions,
respectively.}
\end{figure}

We now turn to our global fit, which uses all available $D^0$, $D^+$, and
$D^{*+}$ data, from Belle, CLEO, ALEPH, and OPAL.
The resulting values of the fit parameters and of $\overline{\chi^2}$ are also
included in Tables~\ref{Tab:D0Fits}--\ref{Tab:D+stFits}.
The $D^0$, $D^+$, and $D^{*+}$ data are compared with the respective
theoretical results based on the global fit in
Figs.~\ref{Fig:D0Global}--\ref{Fig:D+stGlobal}, respectively.
Comparing Figs.~\ref{Fig:D0Global}--\ref{Fig:D+stGlobal} with
Figs.~\ref{Fig:OPALFits} and \ref{Fig:BELLEFits}, we find the most striking
difference to be that the global fit describes the total $H_c$ samples from
ALEPH and OPAL, especially the $D^0$ and $D^+$ samples from OPAL, less well in
the large-$x$ range, for $x\agt0.6$, than the ALEPH/OPAL fits.
This may be understood by observing that, in the global fit, the $c\to H_c$
FFs, which dominantly contribute in the large-$x$ range, are mostly
constrained by the more precise Belle and CLEO data and are appreciably
increased in that $x$ range compared to their counterparts from the ALEPH/OPAL
fits.
On the other hand, the modifications of the $b\to H_c$ FFs are less
significant and do not worsen the agreement with the ALEPH and OPAL data in
any visible way.
The observation that the $c\to D^{*+}$ FF fitted to Belle and CLEO data does
not yield a satisfactory description of the ALEPH data was also made in
Ref.~\cite{Cacciari:2005uk}, where it was speculatively linked to the presence
of large non-perturbative power corrections. 
New experimental data at intermediate energies, between those of the $B$ and
$Z$ factories, which are not expected to become available in the foreseeable
future, would certainly shed more light on this potential anomaly.

\begin{table}
\caption{\label{Tab:D0Fits-m0}Values of fit parameters for the $D^0$ meson
resulting from the Belle/CLEO, OPAL, and global fits in the ZM approach
together with the values of $\overline{\chi^2}$ achieved.}
\begin{center}
\begin{tabular}{|l|c|c|c|c|} \hline
 & Belle/CLEO-ZM & OPAL-ZM & global-ZM \\ 
\hline
$N_c$ & $1.03 \cdot 10^7$ & $3.43 \cdot 10^4$ & $1.04 \cdot 10^7$ \\
$a_c$ & 1.48 & 1.48 & 1.50 \\
$\gamma_c$ & 3.60 & 2.80 & 3.60 \\
\hline
$N_b$ & 13.4 & 13.4 & 80.8 \\
$a_b$ & 3.96 & 3.96 & 5.77 \\
$\gamma_b$ & 0.923 & 0.923 & 1.15 \\
\hline
$\overline{\chi ^2}$ & 3.25 & 0.789 & 4.66 \\
\hline
\end{tabular}
\end{center}
\end{table}
\begin{table}
\caption{\label{Tab:D+Fits-m0}Values of fit parameters for the $D^+$ meson
resulting from the Belle/CLEO, OPAL, and global fits in the ZM approach
together with the values of $\overline{\chi^2}$ achieved.}
\begin{center}
\begin{tabular}{|l|c|c|c|c|} \hline
 & Belle/CLEO-ZM & OPAL-ZM & global-ZM \\ 
\hline
$N_c$ & $7.30 \cdot 10^5$ & $2.62 \cdot 10^4$ & $7.31 \cdot 10^5$ \\
$a_c$ & 1.12 & 1.48 & 1.13 \\
$\gamma_c$ & 3.43 & 2.91 & 3.43 \\
\hline
$N_b$ & 19.0 & 19.0 & 163 \\
$a_b$ & 4.71 & 4.71 & 6.93 \\
$\gamma_b$ & 1.17 & 1.17 & 1.40 \\
\hline
$\overline{\chi ^2}$ & $1.37$ & $0.507$ & $2.21$ \\
\hline
\end{tabular}
\end{center}
\end{table}
\begin{table}
\caption{\label{Tab:D+stFits-m0}Values of fit parameters for the $D^{*+}$
meson resulting from the Belle/CLEO, ALEPH/OPAL, and global fits in the ZM
approach together with the values of $\overline{\chi^2}$ achieved.}
\begin{center}
\begin{tabular}{|l|c|c|c|c|c|} \hline
 & Belle/CLEO-ZM & ALEPH/OPAL-ZM & global-ZM \\ 
\hline
$N_c$ & $1.05 \cdot 10^7$ & $2.80 \cdot 10^4$ & $1.14 \cdot 10^7$ \\
$a_c$ & 0.929 & 1.33 & 1.03 \\
$\gamma_c$ & 3.82 & 2.93 & 3.82 \\
\hline
$N_b$ & 6.52 & 6.52 & 14.9 \\
$a_b$ & 3.25 & 3.25 & 3.87 \\
$\gamma_b$ & 1.04 & 1.04 & 1.16 \\
\hline
$\overline{\chi ^2}$ & 3.69 & 2.04 & 7.64 \\
\hline
\end{tabular}
\end{center}
\end{table}
In order to study the impact of finite $c$ and $b$ quark masses on our fits,
we repeat them in the ZM approach, where $m_c=m_b=0$, except in the definition
of the starting scale $\mu_0$.
We still have $m_H\ne0$, so that the $x$ distributions have finite lower
endpoints and differ from the corresponding $x_p$ distributions in shape.
The resulting values of the fit parameters and of $\overline{\chi^2}$ are
listed for the $D^0$, $D^+$, and $D^{*+}$ mesons in
Tables~\ref{Tab:D0Fits-m0}--\ref{Tab:D+stFits-m0}, respectively.
Comparing Tables~\ref{Tab:D0Fits-m0}--\ref{Tab:D+stFits-m0} with
Tables~\ref{Tab:D0Fits}--\ref{Tab:D+stFits}, we observe that the inclusion of
finite quark masses reduces the $\overline{\chi^2}$ values of the global fits
by 11--16\% and also tends to reduce the $\overline{\chi^2}$ values of the
Belle/CLEO fits, except for the case of $D^{*+}$ mesons, where the difference
is insignificant.
As expected, the quality of the ALEPH/OPAL fits are practically unaffected by
finite-quark-mass effects, which provides a retrospective justification for
the use of the ZM approach in
Refs.~\cite{Kniehl:2006mw,Binnewies:1997gz,Kniehl:2005de}.

\begin{figure}[ht]
\begin{center}
\begin{tabular}{ll}
\parbox{0.45\textwidth}{
\epsfig{file=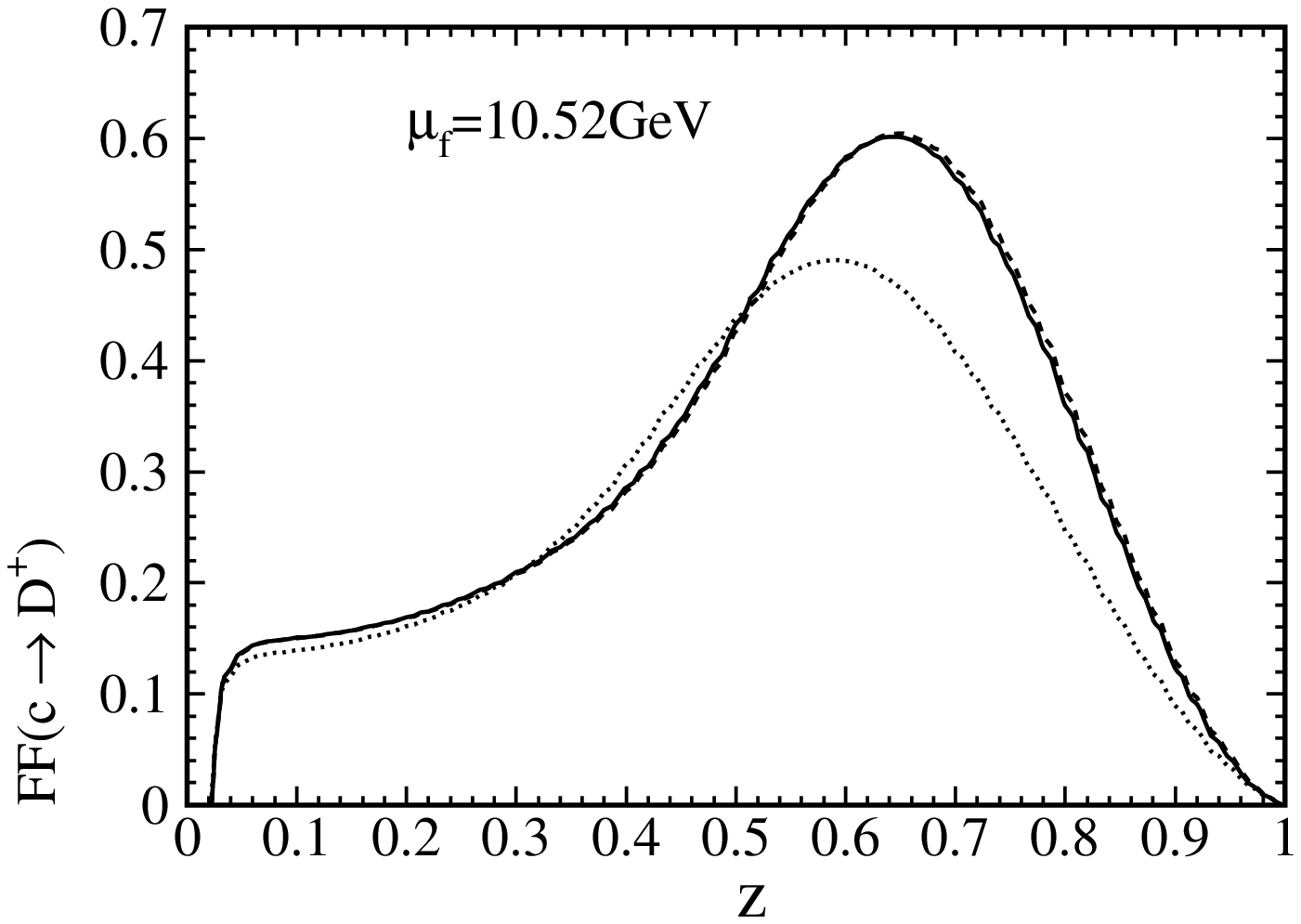,width=0.45\textwidth}
}
&
\parbox{0.45\textwidth}{
\epsfig{file=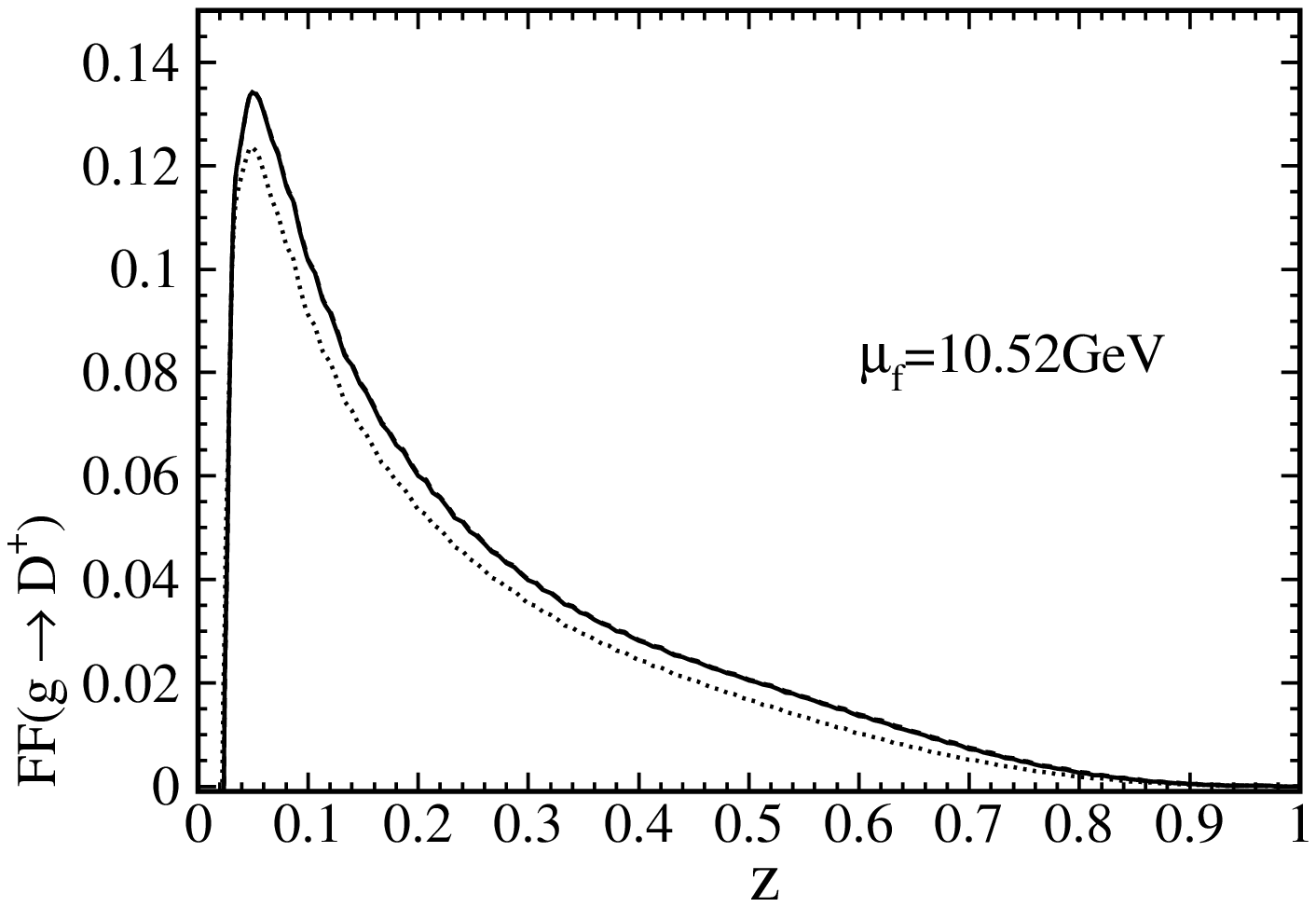,width=0.45\textwidth}
}
\\
(a) & (b)
\end{tabular}
\end{center}
\caption{\label{Fig:D+FFs}(a) $c\to D^+$ and (b) $g\to D^+$ FFs of the
Belle/CLEO fits at $\mu_f=10.52$~GeV as functions of $z$ in the GM approach
(solid lines) and in the ZM approaches with $m_H\ne0$ (dashed lines) and
$m_H=0$ (dotted lines).}
\end{figure}
\begin{figure}[ht]
\begin{center}
\epsfig{file=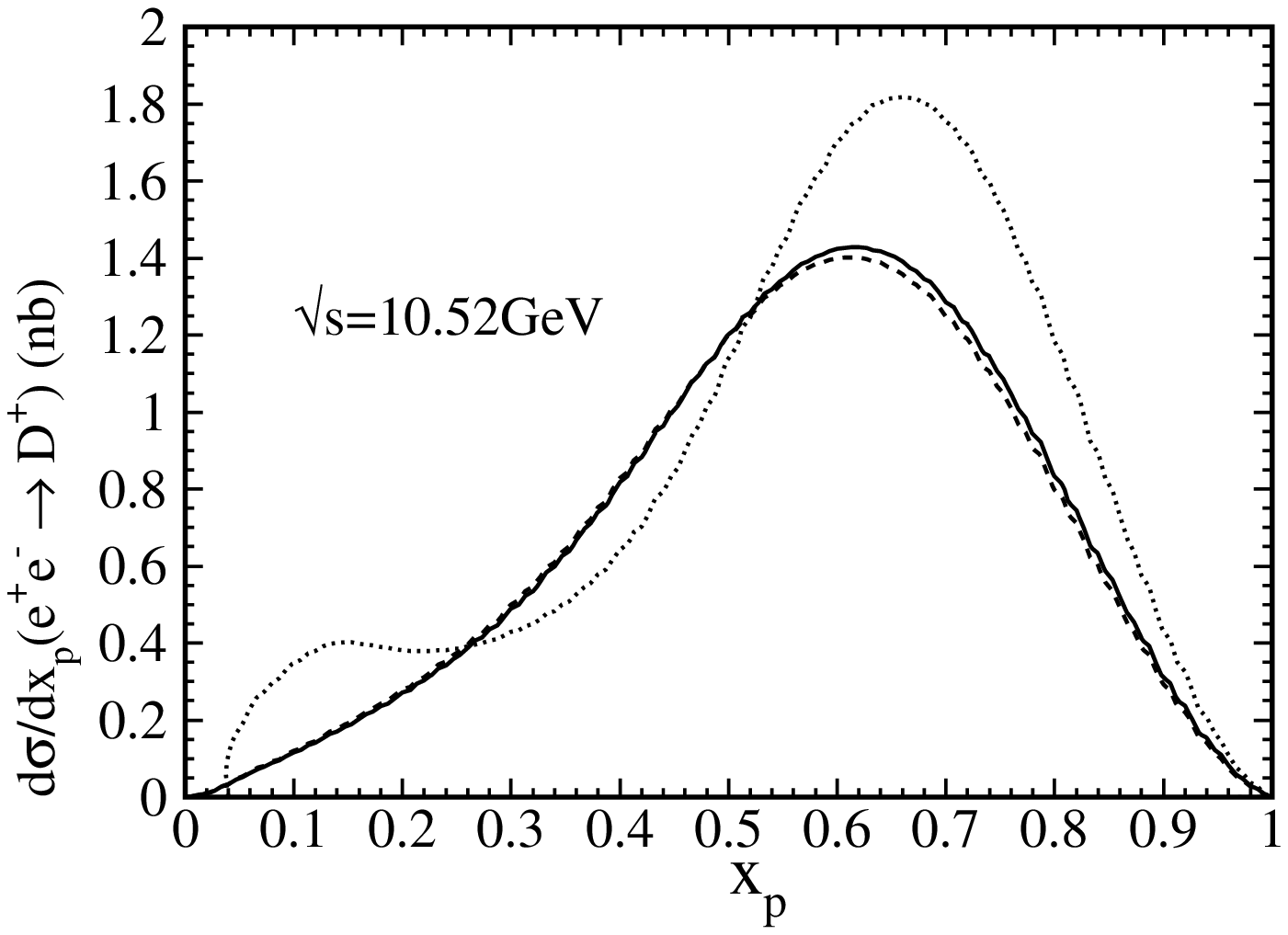,width=0.9\textwidth}
\end{center}
\caption{\label{Fig:comp}$x_p$ distributions of $e^+e^-\to D^++X$ at
$\sqrt s=10.52$~GeV in the GM approach (solid line) and in the ZM approaches
with $m_H\ne0$ (dashed lines) and $m_H=0$ (dotted lines), all evaluated with
the FFs from the Belle/CLEO-GM fit.}
\end{figure}

From the comparison of the fit parameters in the GM and ZM approaches
presented in Tables~\ref{Tab:D0Fits}--\ref{Tab:D+stFits} and
\ref{Tab:D0Fits-m0}--\ref{Tab:D+stFits-m0}, respectively, it is hard to judge
by how much the FFs of the two approaches actually differ as functions of $z$
at a given value of $\mu_f$.
As an example, we thus display in Figs.~\ref{Fig:D+FFs}(a) and (b) the $z$
dependences at $\mu_f=10.52$~GeV of the $c\to D^+$ and $g\to D^+$ FFs,
respectively, of the Belle/CLEO fits in the GM and ZM approaches.
We notice that there is only little difference between the GM and ZM results.
This observation may be understood from Fig.~\ref{Fig:comp}, where the ZM
result for the $x_p$ distribution of
$e^+e^-\to D^++X$ at $\sqrt s=10.52$~GeV evaluated with the GM FFs is compared
with the proper GM result.
In fact, the finite-$m_c$ correction to the hard-scattering cross section only
amounts to a few percent.
A similar observation was made in Ref.~\cite{Nason:1999zj}
using perturbative FFs \cite{Mele:1990cw}.

In the above implementation of the ZM approach, $m_H$ is identified with its
physical values.
It is interesting to study the impact of the finite-$m_H$ correction.
To this end, we repeat the Belle/CLEO-ZM fit for the $D^+$ meson putting also
$m_H=0$, which implies that $x_p=x$ and
$(\diff\sigma/\diff x_p)(x_p)=(\diff\sigma/\diff x)(x)$, as may be gleaned
from Eq.~(\ref{eq:xp}).
In order to obtain an acceptable value of $\overline{\chi^2}$, we exclude the
six data points with $x_p<0.2$ from the fit.
Furthermore, we require that $\tau>\rho_{D^+}$ in Eq.~(\ref{eq:master}), in
which $\rho_H=0$ is put otherwise, to ensure that the hadronic energy after
ISR is above the production threshold.
The resulting $c\to D^+$ and $g\to D^+$ FFs are also shown in
Figs.~\ref{Fig:D+FFs}(a) and (b), respectively.
We notice that the former significantly differs from its counterpart in the
proper ZM approach, its peak being reduced in size and shifted to a lower
value of $z$, while the latter is only moderately affected.
This modification of the FFs is compensated by a reciprocal change in the
line shape of the $x_p$ distribution, as may be seen from Fig.~\ref{Fig:comp},
which also contains the result of the ZM approach with $m_H=0$ evaluated with
the Belle/CLEO-GM FFs.
In fact, the peak position and height are substantially increased relative to
the evaluations with $m_H\ne0$.
The situation is similar for the $D^0$ and $D^{*+}$ mesons.

\begin{table}
\caption{\label{Tab:Branch}Values of $c\to H_c$ and $b\to H_c$ branching
fractions at $\mu_f=2m_b$, 10.52~GeV, and $m_Z$.}
\begin{center}
\begin{tabular}{|c|c|c|c|c|c|c|} \hline
FF set & $H_c$ & $B_c(10.52~\mbox{GeV})$ & $B_c(m_Z)$ & $B_b(2m_b)$ &
$B_b(m_Z)$ \\
\hline
Belle/CLEO-GM & $D^0$ & 0.525 & 0.611 & 0.146 & 0.492 \\
 & $D^+$ & 0.232 & 0.269 & 0.0590 & 0.168 \\
 & $D^{*+}$ & 0.211 & 0.249 & 0.0696 & 0.206 \\
\hline
ALEPH/OPAL-GM & $D^0$ & 0.493 & 0.591 & 0.146 & 0.491 \\
 & $D^+$ & 0.185 & 0.220 & 0.0590 & 0.167 \\
 & $D^{*+}$ & 0.200 & 0.247 & 0.0695 & 0.206 \\
\hline
global-GM & $D^0$ & 0.522 & 0.608 & 0.140 & 0.490 \\
 & $D^+$ & 0.230 & 0.268 & 0.0512 & 0.157 \\
 & $D^{*+}$ & 0.206 & 0.245 & 0.0716 & 0.212 \\
\hline
Belle/CLEO-ZM & $D^0$ & 0.534 & 0.622 & 0.146 & 0.490 \\
 & $D^+$ & 0.235 & 0.273 & 0.0592 & 0.167 \\
 & $D^{*+}$ & 0.215 & 0.254 & 0.0695 & 0.205 \\
\hline
ALEPH/OPAL-ZM & $D^0$ & 0.489 & 0.587 & 0.146 & 0.489 \\
 & $D^+$ & 0.185 & 0.221 & 0.0591 & 0.166 \\
 & $D^{*+}$ & 0.201 & 0.248 & 0.0694 & 0.204 \\
\hline
global-ZM & $D^0$ & 0.527 & 0.614 & 0.141 & 0.488 \\
 & $D^+$ & 0.234 & 0.272 & 0.0517 & 0.157 \\
 & $D^{*+}$ & 0.209 & 0.248 & 0.0718 & 0.210 \\
\hline
\end{tabular}
\end{center}
\end{table}
Besides the $c\to H_c$ and $b\to H_c$ FFs themselves, also their first two
moments are of phenomenological interest.
They correspond to the branching fractions,
\begin{equation}
B_Q(\mu_f)=\int\limits_{\max(\sqrt{\rho_H},z_{\rm cut})}^1\diff z\,
D_Q(z,\mu_f),
\label{eq:bq}
\end{equation}
where $Q=c,b$, and the average fraction of energy that the $H_c$ meson
receives from the $Q$ quark,
\begin{equation}
\langle z\rangle_Q(\mu_f)=\frac{1}{B_Q(\mu_f)}
\int\limits_{\max(\sqrt{\rho_H},z_{\rm cut})}^1\diff z\,zD_Q(z,\mu_f),
\label{eq:xq}
\end{equation}
where the cut $z_{\rm cut}=0.1$ excludes the problematic $z$ range where our
formalism is not valid.
As may be seen from Figs.~\ref{Fig:OPALFits} and \ref{Fig:BELLEFits}, there
are no experimental data at $z<z_{\rm cut}$ either.
Tables~\ref{Tab:Branch} and \ref{Tab:AverageEnergy} contain the values of
$B_Q(\mu_f)$ and $\langle z\rangle_Q(\mu_f)$, respectively, for $Q=c,b$ and
$H_c=D^0,D^+,D^{*+}$ at $\mu_f=2m_b$, 10.52~GeV, and $m_Z$ for the Belle/CLEO,
ALEPH/OPAL, and global fits in the GM and ZM approaches.
We observe from Table~\ref{Tab:Branch} that, within the GM and ZM approaches,
the values of $B_c(\mu_f)$ from the ALEPH/OPAL fits are somewhat smaller than
those from the Belle/CLEO fits, by less than 10\% for the $D^0$ and $D^{*+}$
mesons and by approximately 20\% for the $D^+$ meson.
The corresponding results from the global fits tend to lie between those
from the Belle/CLEO and ALEPH/OPAL fits, but closer to the former.
The GM and ZM approaches yield very similar results.
As for $B_b(\mu_f)$, the differences between three fits and two approaches are
minor.
We note that the values of $B_b(2m_b)$ have to taken with a grain of salt
because, in conrast to the $c\to H_c$ FFs, the $b\to H_c$ FFs are not directly
constrained by low-energy data.
Looking at Table~\ref{Tab:AverageEnergy}, we see that the values of
$\langle z\rangle_Q(\mu_f)$ are shifted towards smaller values through the
DGLAP evolution in $\mu_f$, as expected, and that quark-mass effects are
insignificant here.
As for $\langle z\rangle_c(\mu_f)$, the values from the ALEPH/OPAL fits fall
5--9\% below those from the Belle/CLEO fits, which are only slightly larger
than those from the global fits.
As for the ALEPH/OPAL and global fits, this trend may already be noticed by
comparing Figs.~\ref{Fig:OPALFits}(a), (b), and (c) with
Figs.~\ref{Fig:D0Global}(b), \ref{Fig:D+Global}(b), and
\ref{Fig:D+stGlobal}(b), respectively.
On the other hand, the differences between the various fits are marginal for
$\langle z\rangle_b(\mu_f)$.
\begin{table}
\caption{\label{Tab:AverageEnergy}Values of average energy fractions for
$c\to H_c$ and $b\to H_c$ transitions at $\mu_f=2m_b$, 10.52~GeV, and $m_Z$.}
\begin{center}
\begin{tabular}{|c|c|c|c|c|c|c|} \hline
FF set & $H_c$ & $\langle z\rangle_c(10.52~\mbox{GeV})$ &
$\langle z\rangle_c(m_Z)$ & $\langle z\rangle_b(2m_b)$ &
$\langle z\rangle_b(m_Z)$ \\
\hline
Belle/CLEO-GM & $D^0$ & 0.623 & 0.479 & 0.470 & 0.273 \\
 & $D^+$ & 0.629 & 0.484 & 0.470 & 0.293 \\
 & $D^{*+}$ & 0.659 & 0.503 & 0.508 & 0.305 \\
\hline
ALEPH/OPAL-GM & $D^0$ & 0.591 & 0.450 & 0.470 & 0.273 \\
 & $D^+$ & 0.596 & 0.455 & 0.470 & 0.293 \\
 & $D^{*+}$ & 0.614 & 0.462 & 0.508 & 0.305 \\
\hline
global-GM & $D^0$ & 0.621 & 0.477 & 0.453 & 0.274 \\
 & $D^+$ & 0.629 & 0.484 & 0.451 & 0.288 \\
 & $D^{*+}$ & 0.655 & 0.499 & 0.501 & 0.306 \\
\hline
Belle/CLEO-ZM & $D^0$ & 0.624 & 0.480 & 0.471 & 0.274 \\
 & $D^+$ & 0.632 & 0.486 & 0.470 & 0.293 \\
 & $D^{*+}$ & 0.661 & 0.504 & 0.509 & 0.306 \\
\hline
ALEPH/OPAL-ZM & $D^0$ & 0.591 & 0.450 & 0.471 & 0.274 \\
 & $D^+$ & 0.596 & 0.455 & 0.470 & 0.294 \\
 & $D^{*+}$ & 0.613 & 0.461 & 0.509 & 0.306 \\
\hline
global-ZM & $D^0$ & 0.623 & 0.479 & 0.454 & 0.275 \\
 & $D^+$ & 0.631 & 0.486 & 0.452 & 0.289 \\
 & $D^{*+}$ & 0.657 & 0.500 & 0.501 & 0.308 \\
\hline
\end{tabular}
\end{center}
\end{table}

Of course, the $Q\to H_c$ FFs and their moments depend on scale, scheme,
order, and implementation issues such as the functional form of the ansatz at
the starting scale $\mu_0$ and the value of $\mu_0$ itself, and thus do not
represent physical observables by themselves.
Nevertheless, comparisons of the quantities $B_Q(\mu_f)$ and
$\langle z\rangle_Q(\mu_f)$ defined in Eqs.~(\ref{eq:bq}) and (\ref{eq:xq}),
respectively, with their experimental counterparts determined from the
measured $x$ distributions are instructive, as they reveal in how far the
$Q\to H_c$ transitions actually dominate the cross section distributions.
Moreover, these quantities stringently characteristise the lineshape in $x$
of the $Q\to H_c$ FFs at a given value of $\mu_f$ and simplify the comparisons
with our previous FF sets \cite{Binnewies:1997gz,Kniehl:2005de} and those to
be introduced by other authors.

\begin{table}
\caption{\label{Tab:Bexp}Values of $c\to H_c$ and $b\to H_c$ branching
fractions extracted by ALEPH \cite{Barate:1999bg} and OPAL
\cite{Alexander:1996wy,Ackerstaff:1997ki} at $\sqrt s=m_Z$ from their measured
cross section distributions.}
\begin{center}
\begin{tabular}{|c|c|c|} \hline
$Q\to H_c$ & ALEPH & OPAL \\
\hline
$c\to D^0$ & $0.559\pm0.022$ & $0.605\pm0.040$ \\
$c\to D^+$ & $0.238\pm0.024$ & $0.235\pm0.032$ \\
$c\to D^{*+}$ & $0.233\pm0.015$ & $0.222\pm0.020$ \\
\hline
$b\to D^{*+}$ & -- & $0.173\pm0.020$ \\
\hline
\end{tabular}
\end{center}
\end{table}
\begin{table}
\caption{\label{Tab:xexp}Values of average energy fractions for $c\to H_c$
transitions extracted by Belle \cite{Seuster:2005tr} and CLEO
\cite{Artuso:2004pj} at $\sqrt s=10.52$~GeV and by ALEPH \cite{Barate:1999bg}
and OPAL \cite{Alexander:1996wy,Ackerstaff:1997ki} at $\sqrt s=m_Z$ from their
measured cross section distributions.
The values for Belle and CLEO are obtained by converting the corresponding
average momentum fractions quoted in
Refs.~\cite{Seuster:2005tr,Artuso:2004pj}, respectively.}
\begin{center}
\begin{tabular}{|c|c|c|c|c|} \hline
$H_c$ & Belle & CLEO & ALEPH & OPAL \\
\hline
$D^0$ & $0.640\pm0.002$ & $0.640\pm0.005$ & -- & $0.487\pm0.014$ \\
$D^+$ & $0.647\pm0.001$ & $0.650\pm0.007$ & -- & $0.483\pm0.019$ \\
$D^{*+}$ & $0.682\pm0.001$ & $0.682\pm0.006$ & $0.488\pm0.008$ & 
$0.515\pm0.009$ \\
\hline
\end{tabular}
\end{center}
\end{table}
The values of the branching fractions and average energy fractions of the 
$Q\to H_c$ transitions measured by Belle and CLEO at $\sqrt s=10.52$~GeV and
by ALEPH and OPAL at $\sqrt s=m_Z$ are collected in Tables~\ref{Tab:Bexp} and
\ref{Tab:xexp}, respectively.
For comparison, we present in Table~\ref{Tab:phys} the counterparts of
$B_c(m_Z)$, $B_b(m_Z)$, $\langle z\rangle_c(10.52~\mbox{GeV})$, and
$\langle z\rangle_c(m_Z)$ extracted from the cross section distributions based
on the global fits in the GM approach, which are shown in
Figs.~\ref{Fig:D0Global}--\ref{Fig:D+stGlobal}.
\begin{table}
\caption{\label{Tab:phys}Counterparts of $B_c(m_Z)$, $B_b(m_Z)$,
$\langle z\rangle_c(10.52~\mbox{GeV})$, and $\langle z\rangle_c(m_Z)$
extracted from the cross section distributions based on the global fits in the
GM approach.}
\begin{center}
\begin{tabular}{|c|c|c|c|c|} \hline
$H_c$ & ``$B_c(m_Z)$'' & ``$B_b(m_Z)$'' & 
``$\langle z\rangle_c(10.52~\mbox{GeV})$'' & 
``$\langle z\rangle_c(m_Z)$'' \\
\hline
$D^0$ & 0.628 & 0.515 & 0.632 & 0.509 \\
$D^+$ & 0.276 & 0.164 & 0.640 & 0.516 \\
$D^{*+}$ & 0.252 & 0.221 & 0.666 & 0.532 \\
\hline
\end{tabular}
\end{center}
\end{table}

\begin{figure}[ht]
\begin{center}
\begin{tabular}{ll}
\parbox{0.45\textwidth}{
\epsfig{file=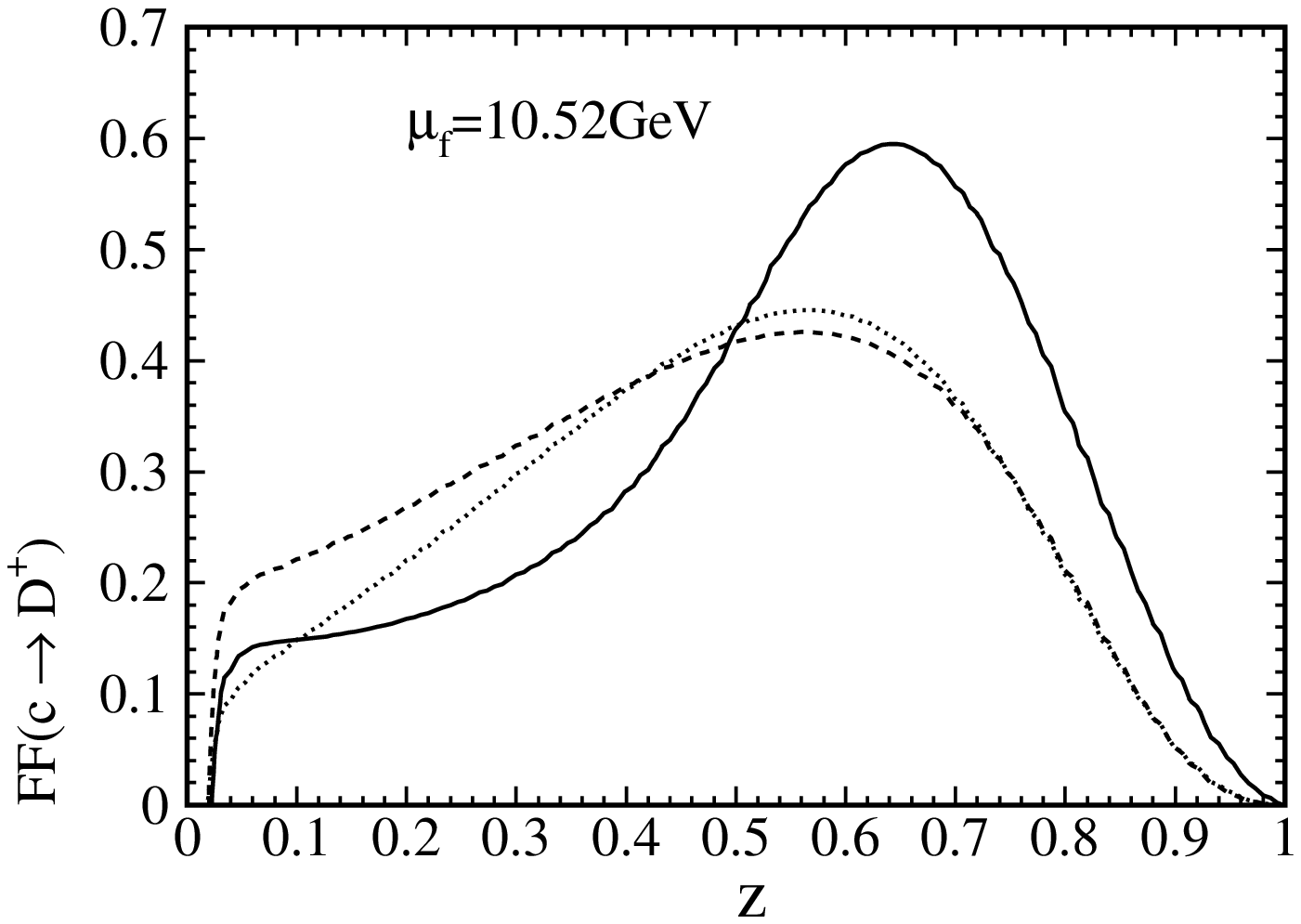,width=0.45\textwidth}
}
&
\parbox{0.45\textwidth}{
\epsfig{file=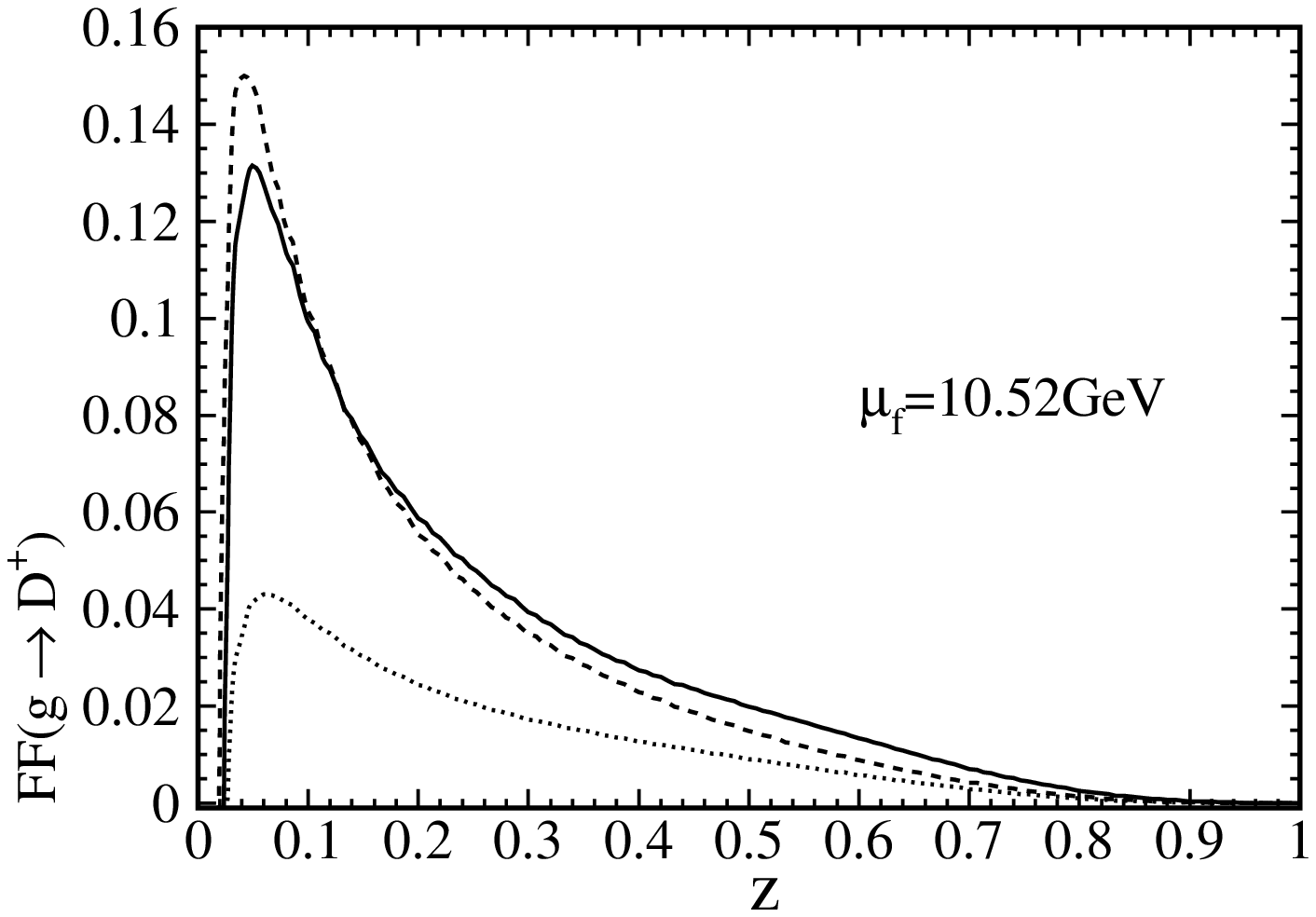,width=0.45\textwidth}
}
\\
(a) & (b)
\end{tabular}
\end{center}
\caption{\label{Fig:old}(a) $c\to D^+$ and (b) $g\to D^+$ FFs of the global
fit in the GM approach at $\mu_f=10.52$~GeV (solid lines) compared with their
counterparts from Refs.~\cite{Kniehl:2006mw} (dashed lines) and
\cite{Kniehl:2005de} (dotted lines).}
\end{figure}

In the remainder of this section, we compare our favourable FFs, from the
global fit in the GM approach, with those from
Refs.~\cite{Kniehl:2006mw,Binnewies:1997gz,Kniehl:2005de}, which were
determined through fits to ALEPH \cite{Barate:1999bg} and OPAL
\cite{Alexander:1996wy,Ackerstaff:1997ki} data in the ZM approach
parameterizing the $c\to H_c$ and $b\to H_c$ FFs using the Peterson
\cite{Peterson:1982ak} and Kartvelishvili-Likhoded
\cite{Kartvelishvili:1985ac} ansaetze, respectively.
In Ref.~\cite{Kniehl:2006mw}, the initial scale for the DGLAP evolution was
taken to be $\mu_0=m_c,m_b$ as in the present paper, while it was chosen as
$\mu_0=2m_c,2m_b$ in Refs.~\cite{Binnewies:1997gz,Kniehl:2005de}.
As in Figs.~\ref{Fig:D+FFs}(a) and (b), we consider the $c\to D^+$ and
$g\to D^+$ FFs at $\mu_f=10.52$~GeV.
The comparison is presented in Figs.~\ref{Fig:old}(a) and (b).
From Fig.~\ref{Fig:old}(a), we observe that our global-GM $c\to D^+$ FF
significantly differs from those of Refs.~\cite{Kniehl:2006mw,Kniehl:2005de}
in lineshape, which essentially reflects the strong pull of the Belle and CLEO
data and the difference between the Bowler and Peterson parameterisations.
From Fig.~\ref{Fig:old}(b), we see that our global-GM $g\to D^+$ FF is similar
to the one of Ref.~\cite{Kniehl:2006mw}, while the one of
Ref.~\cite{Kniehl:2005de} is strongly suppressed, especially in the lower $z$
range.
As explained in Refs.~\cite{Kniehl:2006mw,Kniehl:2005ej}, this may be
attributed to the reduced length of the evolution path.
The situation is similar for the $D^0$ and $D^{*+}$ mesons.

\section{Conclusions}
\label{sec:four}

Previous determinations of non-perturbative charmed-hadron FFs in the parton
model of QCD
\cite{Kniehl:2006mw,Binnewies:1997gz,Kniehl:2005de} were
based on data from the $Z$-boson resonance, so that the effects of finite
quark and hadron masses were greatly suppressed and could safely be neglected.
The advent of precise data from the $B$ factories offers us the opportunity to
further constrain the charmed-hadron FFs and to test their scaling violations.
However, this 
motivates the incorporation of quark and hadron mass effects,
which are then likely to be no longer negligible, into the formalism.
The GM variable-flavor-number scheme, which we previously advocated
\cite{Kniehl:2005ej,Kniehl:2004fy}, provides a rigorous theoretical framework
for this and is employed here for the first time to determine FFs of heavy
hadrons.

Specifically, we determined here new FFs for $D^0$, $D^+$, and $D^{*+}$ mesons
through global fits to all available $e^+e^-$ annihilation data, from Belle
\cite{Seuster:2005tr}, CLEO \cite{Artuso:2004pj}, ALEPH \cite{Barate:1999bg},
and OPAL \cite{Alexander:1996wy,Ackerstaff:1997ki}.
In contrast to the situation at the $Z$-boson resonance, the $x$ distribution
of the cross section for continuum production is appreciably distorted by the
effects of electromagnetic ISR, which we, therefore, took into account.
For comparison, we also performed fits to individual data sets.
We found that the global fits somewhat suffer from the fact that the Belle and
CLEO data tend to drive the average $x$ value of the $c\to H_c$ FFs to larger
values, which leads to a worse description of the ALEPH and OPAL data.
Since the $b\to H_c$ FFs are only indirectly constrained by the Belle and CLEO
data, their form is only feebly affected by the inclusion of these data in the
fits.
In order to assess the significance of finite-mass effects, we repeated the
fits in the ZM variable-flavor-number scheme.
As expected, the inclusion of finite-mass effects tends to improve the overall
description of the data, by reducing the $\overline{\chi^2}$ values achieved.
Specifically, hadron mass effects turned out to be more important than quark
mass effects.
In fact, they are indispensable to usefully describe the low-$x_p$ tails of
the measured cross sections.

A {\tt FORTRAN} subroutine that evaluates the FFs presented here for given
values of $z$ and $\mu_f$ may be obtained from the authors upon request via
electronic mail.

\section*{Acknowledgment}

The work of T.K., B.A.K., and G.K. was supported in part by the German Federal
Ministry for Education and Research BMBF through Grant No.\ 05~HT6GUA and by
the German Research Foundation DFG through Grant No.\ KN~365/7--1.
The work of T.K. was also supported in part by the DFG through Graduate
School No.\ GRK~602 {\it Future Developments in Particle Physics}.

\appendix

\section{Effective electroweak charges}
\label{sec:appa}

The effective electroweak charges appearing in Eqs.~(\ref{eq:ZM-XS}),
(\ref{eq:sigtot}), (\ref{eq:gm}), and (\ref{eq:gmg}) are given by
\begin{eqnarray}
V_{q_i}^2 &=& e_e^2 e_{q_i}^2 + 2 e_e v_e e_{q_i} v_{q_i} \rho_1(s) 
+ \left(v_e^2+a_e^2\right)v_{q_i}^2 \rho_2(s),
\nonumber\\
A_{q_i}^2 &=& \left(v_e^2+a_e^2\right)a_{q_i}^2 \rho_2(s),
\end{eqnarray}
where $v_f=(T_{3f}-2e_f\sin^2\theta_w)/(2\sin\theta_w\cos\theta_w)$ and
$a_f=T_{3f}/(2\sin\theta_w\cos\theta_w)$ are the vector and axial-vector
couplings of fermion $f$, with fractional electric charge $e_f$ and third
component $T_{3f}$ of weak isospin, to the $Z$ boson, and
\begin{eqnarray}
\rho_1(s) &=& \frac{s(s-m_Z^2)}{(s-m_Z^2)^2+m_Z^2\Gamma_Z^2},
\nonumber\\
\rho_2(s) &=& \frac{s^2}{(s-m_Z^2)^2+m_Z^2\Gamma_Z^2}
\end{eqnarray}
are propagator functions.
Here, $\theta_w$ is the weak mixing angle and $\Gamma_Z$ is the total decay
width of the $Z$ boson.
For small energies, $\sqrt{s}\ll m_Z$, the propagator functions $\rho_1(s)$
and $\rho_2(s)$ are negligible.

\boldmath
\section{Single heavy-quark inclusive cross sections at
$\mathcal{O}(\alpha_s)$}
\label{sec:appb}
\unboldmath

In this appendix, we list the NLO coefficient functions appearing in
Eqs.~(\ref{eq:XS-NLODecomposition}) and (\ref{eq:gmg}).
We cast our results in a form similar to Ref.~\cite{Nason:1993xx}, except that
our formulas include the overall factor $C_F$.
We start by introducing the short-hand notation:
\begin{eqnarray}
\tau_x & = & 1-x,
\nonumber\\
\beta_x&=&\sqrt{1-\frac{\rho}{\tau_x}},
\nonumber\\
\xi(x,\rho) & = & \ln\frac{\rho-2x-2\sqrt{x^2-\rho}}{\rho-2x+2\sqrt{x^2-\rho}}.
\end{eqnarray}
The coefficient functions due to virtual-soft corrections to inclusive
single heavy-quark production read:
\begin{eqnarray}
S^{(v)}_T(\rho) & = & \frac{C_F}{2} \left\{(2-\rho)\left[ 4 \ln \frac{4}{\rho}
\ln \frac{1+\beta}{1-\beta} - 4 \li\left( -\frac{1-\beta}{2\beta} \right) 
- 2\ln ^2 \frac{2\beta}{1-\beta}\right.\right. \nonumber\\
&&{} +\left.\frac{4}{3} \pi^2 + \ln^2 \frac{1+\beta}{1-\beta} 
+ \li \left( -\frac{4\beta}{(1-\beta)^2}\right) 
- \li\left(\frac{4\beta}{(1+\beta)^2}\right)\right] \nonumber\\
&& +\left.(10-8\rho) \ln \frac{1+\beta}{1-\beta} -4\beta 
-8\beta \ln\frac{4}{\rho} \right\},
\nonumber\\
S^{(v)}_L(\rho) & = & \frac{\rho}{2}S^{(v)}_T(\rho) 
-  C_F\frac{\rho \beta^2}{2} \ln \frac{1+\beta}{1-\beta},
\nonumber\\
S^{(a)}_T(\rho) & = & \beta^2 S^{(v)}_T(\rho) 
+ 2C_F \rho \beta^2 \ln \frac{1+\beta}{1-\beta},
\nonumber\\
S^{(a)}_L(\rho) & = & 0.
\end{eqnarray}
The coefficient functions due to real corrections to inclusive single
heavy-quark production read:
\begin{eqnarray}
R^{(v)}_T(x,\rho) & = & C_F\left\{
\frac{2}{\sqrt{x^2-\rho}}\left[ \rho(2-\tau^2_x) 
+ 4\frac{\tau^2_x(1+\tau_x)^3}{(4\tau_x+\rho)^2} + \tau_x (4+\tau_x) 
\left( 1-\frac{2\tau_x(1+\tau_x)}{4\tau_x+\rho} \right)-2\right]\right.
\nonumber\\
&&{}+\left. \frac{\rho^2(2-\tau_x^2)+\rho(2x^3-7x^2-1)+2x^2(1+x^2)}
{2(x^2-\rho)}\xi(x,\rho)\right\},
\nonumber\\
R^{(v)}_L(x,\rho) & = &C_F\left\{
\frac{2}{\sqrt{x^2-\rho}}\left[ -\rho(1-\rho)-\tau_x
(\tau_x-2\rho) + \frac{2\tau_x^2(1+\tau_x)}{4\tau_x+\rho}\right]\right.
\nonumber\\
&&{} +\left.
\frac{\rho^3+\rho^2(4\tau_x-3)+\rho(3x^2-1)}{2(x^2-\rho)}\xi(x,\rho)\right\},
\nonumber\\
R^{(a)}_T(x,\rho) & = & C_F\left\{
\frac{4}{\sqrt{x^2-\rho}} \left[ -\rho^2 +2\rho x
+\tau^3_x +\frac{3}{2} \tau^2_x +2\tau_x-1
+\frac{2\tau_x^2(1+\tau_x)^3}{(4\tau_x+\rho)^2}
\right.\right.
\nonumber\\
&&{} -\left.\frac{\tau_x^2(1+\tau_x)(5\tau_x +4)}{4\tau_x+\rho} \right]
\nonumber\\
&&{} +\left. \frac{-2\rho^3+8\rho^2 x+\rho x^2 (2\tau_x-9) - \rho + 2x^2(1+x^2)}
{2(x^2-\rho)} \xi(x,\rho)\right\},
\nonumber\\
R^{(a)}_L(x,\rho) & = &C_F\left\{
\frac{2\tau_x^2}{\sqrt{x^2-\rho}} \left[ \rho+x^2-5
-\frac{8\tau_x(1+\tau_x)^3}{(4\tau_x+\rho)^2} - \frac{2(1+\tau_x)}
{4\tau_x+\rho}(\tau_x^2-8\tau_x-2)\right]\right.
\nonumber\\
&&{} +\left.\frac{\tau_x^2 \rho (\rho+\tau_x^2 -2)}{2(x^2-\rho)} \xi(x,\rho)
\right\}.
\end{eqnarray}
The coefficient functions due to real corrections to inclusive single gluon
production read:
\begin{eqnarray}
G_T^{(v)}(x,\rho) & = &C_F\left\{
2\left[ \frac{1+(1-x)^2}{x} + \rho \frac{1-x}{x} 
- \frac{\rho^2}{2x} \right] \left( \ln \frac{1+\beta_x}{1-\beta_x} 
-\beta_x \right)\right.
\nonumber\\
&&{} -\left.4\frac{1-x}{x}\beta_x - \frac{\rho^2 \beta_x}{x}\right\},
\nonumber\\
G_L^{(v)}(x,\rho) & = & C_F \left[-\frac{2\rho}{x}
\ln\frac{1+\beta_x}{1-\beta_x} + 4\beta_x \frac{1-x}{x}\right],
\nonumber\\
G_T^{(a)}(x,\rho) & = &C_F\left\{
2\left[\frac{1+(1-x)^2}{x} - \rho \frac{1-x}{x} 
+ \frac{\rho^2}{2x} \right] \left( \ln \frac{1+\beta_x}{1-\beta_x} 
-\beta_x \right)\right.
\nonumber\\
&&{} -\left.4\frac{1-x}{x}\beta_x + \frac{\rho^2 \beta_x}{x}\right\},
\nonumber\\
G_L^{(a)}(x,\rho) & = & C_F \left[ \rho \frac{\rho+x^2+2x-4}{x}
\ln \frac{1+\beta_x}{1-\beta_x} + 2\beta_x (2+\rho)\frac{1-x}{x} \right].
\end{eqnarray}

In order to establish the subtraction terms to be included in the GM result to
ensure matching with the ZM result in the massless limit, we need to take the
limit $m\to0$ in the virtual-soft and real correction terms listed above.
In this limit, the vector and axial-vector parts coincide.
Specifically, we have
\begin{eqnarray}
\lim\limits_{\rho\to0}\left[S_T^{(u)}(\rho)+S_L^{(u)}(\rho)\right]&=&
C_F\left(\ln^2\frac{4}{\rho}+\ln\frac{4}{\rho}-2+\pi^2\right),
\nonumber\\
\lim\limits_{\rho\to0}\frac{R_T^{(u)}(x,\rho)+R_L^{(u)}(x,\rho)}{(1-x)_+}
&=&C_F\left\{\delta(1-x)\left(-\ln^2\frac{4}{\rho}
+\frac{1}{2}\ln\frac{4}{\rho}-\frac{1}{2}-\frac{\pi^2}{3}\right)\right.
\nonumber\\
&&+{}\left(\frac{1}{1-x}\right)_+\left[(1+x^2)\ln\frac{4}{\rho}
-4x+\frac{x^2}{2}\right]
\nonumber\\
&&{}-\left.(1+x^2)\left[\frac{\ln(1-x)}{1-x}\right]_+
+2\frac{1+x^2}{1-x}\ln x\right\},
\nonumber\\
\lim\limits_{\rho\to0}\left[G_T^{(u)}(x,\rho)+G_L^{(u)}(x,\rho)\right]&=&
2C_F\frac{1+(1-x)^2}{x}\left[\ln\frac{4}{\rho}+\ln(1-x)-1\right].
\end{eqnarray}
Comparison of Eqs.~(\ref{eq:gm}) and (\ref{eq:gmg}) with Eq.~(\ref{eq:ZM-XS})
yields
\begin{eqnarray}
\lim\limits_{\rho\to0}\left\{
\delta(1-x)\left[S_T^{(u)}(\rho)+S_L^{(u)}(\rho)\right]
+\frac{R_T^{(u)}(x,\rho)+R_L^{(u)}(x,\rho)}{(1-x)_+}\right\}&=&
P_{q\to q}^{(0,T)}(x)\ln\frac{s}{\mu_f^2}+C_q(x)
\nonumber\\
&&{}+d_q^{(1)}(x,\mu_f),
\nonumber\\
\lim\limits_{\rho\to0}\left[G_T^{(u)}(x,\rho)+G_L^{(u)}(x,\rho)\right]
&=&2\left[P_{q\to g}^{(0,T)}(x)\ln\frac{s}{\mu_f^2}+C_g(x)\right]
\nonumber\\
&&{}+d_g^{(1)}(x,\mu_f),
\end{eqnarray}
where \cite{Nason:1993xx,Mele:1990cw}
\begin{eqnarray}
d_q^{(1)}(x,\mu_f)&=&P_{q\to q}^{(0,T)}(x)\ln\frac{\mu_f^2}{m^2}
+C_F(1+x^2)\left\{\delta(1-x)-\left(\frac{1}{1-x}\right)_+
-2\left[\frac{\ln(1-x)}{1-x}\right]_+\right\},
\nonumber\\
d_g^{(1)}(x,\mu_f)&=&2P_{q\to g}^{(0,T)}(x)\ln\frac{\mu_f^2}{m^2}
-2C_F(2\ln x+1).
\end{eqnarray}


\end{document}